\documentclass[twocolumn,pra,notitlepage]{revtex4-1}
\usepackage{amsmath}
\usepackage{graphicx}
\usepackage{wrapfig}
\usepackage{enumerate}
\usepackage{float}

\begin{document}

\title{Collimation and finite-size effects in suspended resonant guided-mode gratings}

\author{Christian Toft-Vandborg, Alexios Parthenopoulos, Ali Akbar Darki, and Aur\'{e}lien Dantan}\email{dantan@phys.au.dk}

\address{Department of Physics and Astronomy, Aarhus University, DK-8000 Aarhus C, Denmark}


\begin{abstract}
The optical transmission of resonant guided-mode gratings patterned on suspended silicon nitride thin films and illuminated at normal incidence with a Gaussian beam are investigated both experimentally and theoretically. Effects due to the beam focusing and its finite size are accounted for by a phenomenological coupled mode model whose predictions are found to be in very good agreement with the experimentally measured spectra for various grating structures and beam sizes, and which allow for a detailed analysis of the respective magnitude of these effects. These results are highly relevant for the design and optimization of such suspended structured films which are widely used for photonics, sensing and optomechanics applications.
\end{abstract}

\date{\today}

\maketitle


\section{Introduction}

Resonant guided-mode structures, such as gratings and photonic crystals, are ubiquitous in photonics and sensing applications~\cite{Wang1993,ChangHasnain2012,Quaranta2018,Cheben2018}. The transverse subwavelength structuring of thin films allows for instance for the realization of a wide variety of ultracompact optical components, such as optical filters, couplers, reflectors, lenses, polarizers, spatial differentiators, lasers, etc. The tailored optical properties of such nanostructured thin films also make them highly suitable for sensing applications, e.g. for bioimaging or environmental sensing~\cite{Wang1993,ChangHasnain2012,Quaranta2018,Cheben2018}. Pretensioned and suspended nanostructured ultrathin films~\cite{Kemiktarak2012,Bui2012,Norte2016,Reinhardt2016,Chen2017,Moura2018} are also attractive for optomechanics applications in which the combination of the films' high mechanical and optical quality can be exploited for photonics and sensing applications.

In this work we focus on canonical one-dimensional subwavelength dieletric grating (SWG) structures, in which the interaction of light impinging on the grating with guided modes in the structure leads to the appearance of Fano resonances for specific wavelengths and polarizations of the incident light. While the resonant interference processes are well-understood and accurately predicted for incident plane waves and infinite periodic structures~\cite{Wang1993,Rosenblatt1997}, it is well-known that finite-size effects due to the finite extent of the structure and/or the illuminating beam, as well as angular effects due to the focusing of the illuminating beam, may strongly affect these interferences and thus limit the optical performances of these structures~\cite{Magnusson1993,Saarinen1995,Brazas1995,Rosenblatt1997,Loktev1997,Glasberg1998,Boye2000,Jacob2000,Jacob2001,Bendickson2001,Thurman2003,Bonnet2003,Tishchenko2003,Peters2004,Tishchenko2004,Niederer2004,Kenyon2008,Ren2015}. Indeed, on the one hand, the finite extent of the incoming beam limits the interaction of the incoming light with the guided modes in the grating. On the other hand, a focused illuminating beam can be regarded as a superposition of plane waves impinging on the grating with a given angular distribution corresponding to a given guided mode resonance distributions. Both effects may thus strongly modify the position and width of the observed Fano resonances, and have been investigated analytically, numerically and experimentally for a number resonant grating and waveguide structures operating in various regimes of coupling strength, incidence, etc.~\cite{Magnusson1993,Saarinen1995,Brazas1995,Rosenblatt1997,Loktev1997,Glasberg1998,Boye2000,Jacob2000,Jacob2001,Bendickson2001,Thurman2003,Bonnet2003,Tishchenko2003,Peters2004,Tishchenko2004,Niederer2004,Kenyon2008,Ren2015}.

While an important motivation of early investigations was the estimation of the achievable resonance linewidths of guided-mode resonant filters, precisely elucidating and quantifying these effects in a simple manner is still highly relevant for a number of current applications of these gratings, e.g., for improving the performances of resonant grating-based optical spatial differentiators~\cite{Bykov2018,Dong2018,Yang2020,Parthenopoulos2021,Cheng2021}, for investigating cavity optomechanics with single~\cite{Kemiktarak2012a,Xu2017} or multiple~\cite{Xuereb2012,Xuereb2014,Piergentili2018,Gartner2018,Wei2019,Manjeshwar2020,Yang2020b} high-reflectivity membranes in optical resonators or for investigating lasing~\cite{Yang2015,Guillemot2020} and optomechanical~\cite{Kemiktarak2014,Naesby2018,Cernotik2019,Fitzgerald2021} phenomena in resonant grating-based microcavities. 

\begin{figure}
\includegraphics[width=0.8\columnwidth]{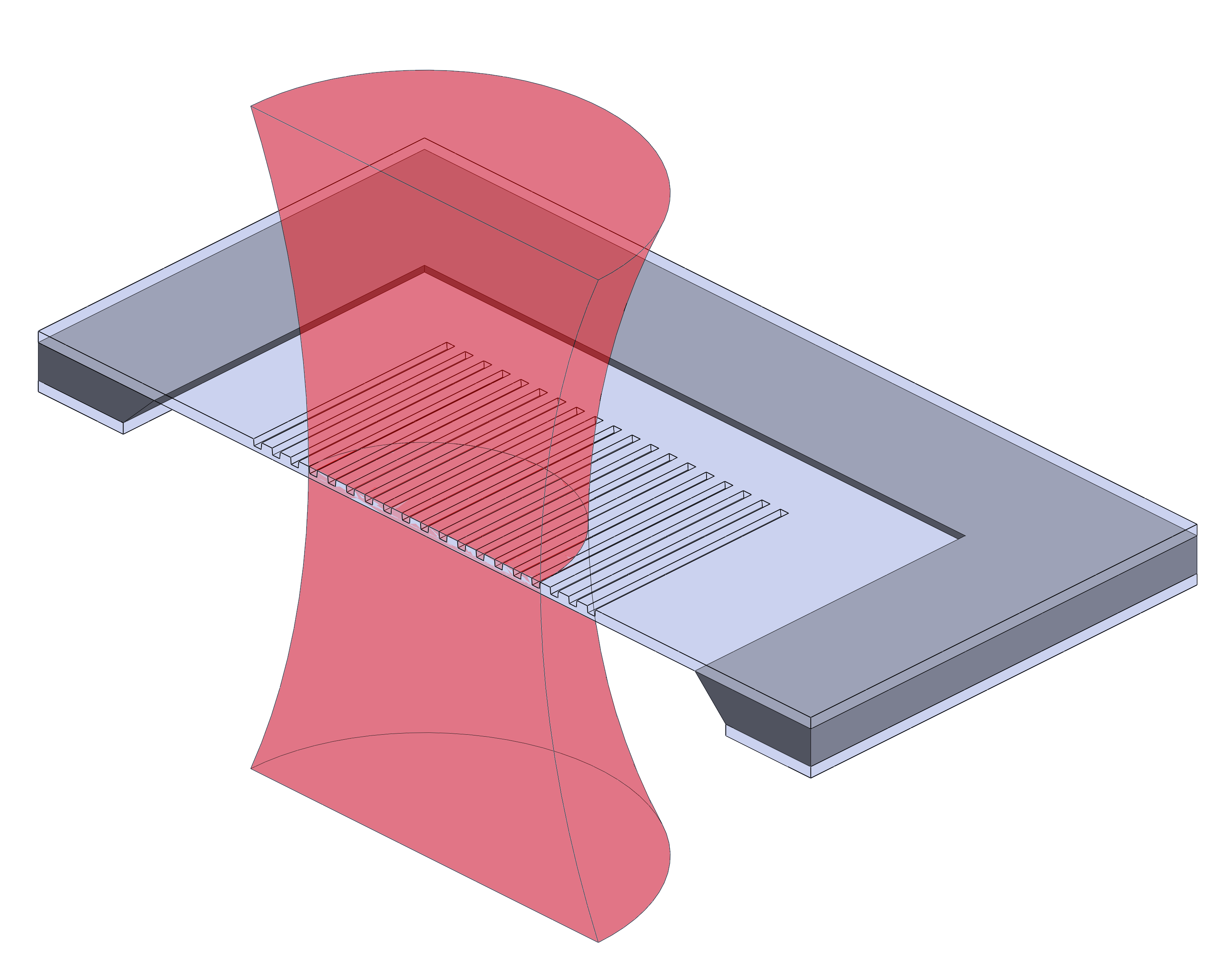}
\caption{Illustration of the situation considered: a suspended thin film patterned with a subwavelength grating is illuminated at normal incidence by a Gaussian beam of linearly polarized, monochromatic light focused on the grating. The beam collimation and finite size affect both the resonant interaction of the incident light with a guided mode in the grating.}
\label{fig:suspended}
\end{figure}

In this article we investigate both experimentally and theoretically finite-size and collimation effects for various resonant guided-mode gratings patterned on suspended, absorption-free dielectric (Si$_3$N$_4$) thin films illuminated at normal incidence by a Gaussian beam. Effects due to the beam focusing and its finite size are both accounted for by a coupled mode model~\cite{Bykov2015}, which we phenomenologically modify to include finite-size effects following a waveguide interference approach put forward by Jacob et al.~\cite{Jacob2000,Jacob2001}. The predictions of this simple analytical model are found to be in very good agreement with the experimentally measured spectra for various grating structures and beam sizes, and allow for a straightforward evaluation of the respective magnitude of these effects. These findings are thus relevant for the design and optimization of suspended structured thin films for the abovementioned applications. Let us additionally note that the results discussed here in a canonical one-dimensional geometry would be relevant for---and could be extended to---two-dimensional photonic crystal structures~\cite{Fan2002,Fan2003,Crozier2006,Grepstad2013,Bernard2016}.

The paper is organized as follows: first, we start by discussing the fabrication and characterization of the suspended resonant guided-mode gratings in Sec.~\ref{sec:fab}, before presenting the optical transmission measurements of five different gratings for various input beam sizes in Sec.~\ref{sec:exp_results}. In Sec.~\ref{sec:model} the theoretical model used for analyzing the experimental spectra is introduced. Section~\ref{sec:analysis} reports on the detailed analysis of both finite-size and collimation effects using the model for the five samples investigated in this work. We conclude in Sec.~\ref{sec:conclusion}.

\section{Experimental methods and results}

\subsection{Sample fabrication and characterization}\label{sec:fab}

The gratings are fabricated following the recipe detailed in~\cite{Nair2019,Parthenopoulos2021} and structurally characterized using AFM profilometry~\cite{Darki2021}. In brief, commercial (Norcada Inc., Canada), high tensile stress ($\sim$GPa) stoichiometric silicon nitride films suspended on a silicon frame are patterned with subwavelength grating structures using Electron Beam Lithography and dry etching. The silicon nitride films used in this work are 203 nm thick and suspended on a 5 mm-square, 200 $\mu$m-square silicon frame. The lateral dimension of the suspended films is 500 $\mu$m, and the lateral size $b$ of the square area patterned with a SWG varies between 100 and 200 $\mu$m. The grating fingers are trapezoidal with a height and wall angle that depend on the etching parameters. The gratings' topology is noninvasively characterized by Atomic Force Microscopy (AFM) profilometry, as described in~\cite{Darki2021}. The film thickness (203 nm) and refractive index (2.0) are determined independently by ellipsometry~\cite{Nair2017}. The gratings' period and duty cycle are chosen so as to observe high reflectivity guided mode resonances in the 915-975 nm range provided by the tunable laser used for their optical characterization.  The geometrical parameters of the five SWGs used in this work, as well as the incident light polarization used for the transmission measurements, are given in Table~\ref{tab:tab}.

\begin{table}[h]
\caption{\label{tab:parameters}Geometrical parameters of the SWGs and incident light polarization.}
\label{tab:tab}
\begin{ruledtabular}
\begin{tabular}{ccccccc}
Sample & b [$\mu$m] & $\Lambda$ [nm] & $w_t$ [nm] & $w_m$ [nm] & $h$ [nm] & Polarization\\
A & 200 & 848 & 340 & 355 & 114 & TM\\
B & 200 & 858 & 394 & 411 & 153 & TM\\
C & 200 & 801 & 416 & 434 & 90 & TM\\
D & 100 & 800 & 432 & 453 & 91 & TM\\
E & 200 & 646 & 474 & 474 & 88 & TE
\end{tabular}
\end{ruledtabular}
\end{table}

\begin{figure}
\includegraphics[width=\columnwidth]{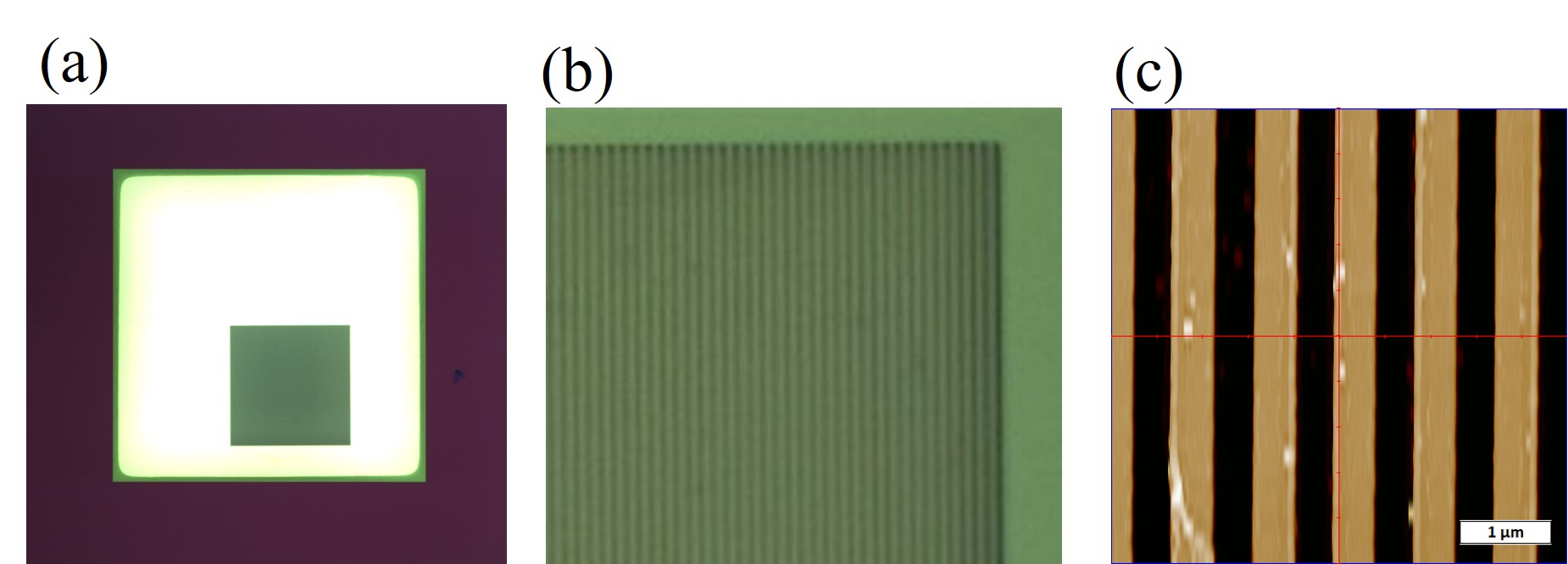}\\
\includegraphics[width=\columnwidth]{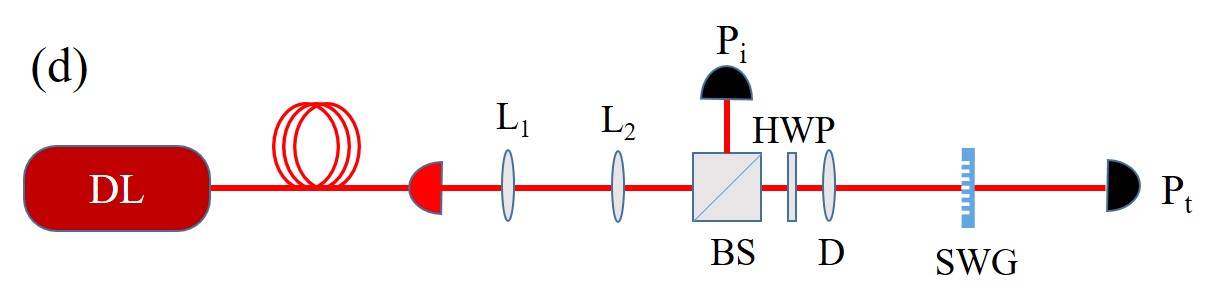}
\caption{(a) Microscope topview picture of sample C showing a 200 $\mu$m-square SWG (green) patterned on a 500 $\mu$m-square suspended Si$_3$N$_4$ film (white). (b) Zoom-in on the top-right corner of the patterned area. (c) Result of an AFM scan of the SWG. (d) Schematic of the setup used for the optical characterization of the SWGs (see text for details). DL: diode laser. L$_1$, L$_2$: lenses. BS: 50:50 beamsplitter. HWP: half-wave plate. D: achromatic doublet. P$_i$, P$_t$: photodiodes.}
\label{fig:setup}
\end{figure}

Figure~\ref{fig:setup} shows the setup used for the optical characterization of the SWGs. Monochromatic light issued from a tunable external cavity diode laser (Toptica DLC Pro) is coupled into a single-mode fiber and its output focused onto the sample using an achromatic 75 mm-focal length doublet $D$ positioned after a 50:50 beamsplitter (BS). The size of the beam at the sample position is adjusted by adjusting the telescope formed lenses $L_1$ and $L_2$. For each beamsize the position and size of the waist is determined by measuring the transmission through a 25 $\mu$m or 50 $\mu$m pinhole. The sample, mounted on a 5-axis translation stage, is positioned at the waist position and such that the light impinges on the SWG at normal incidence. The light polarization is adjusted by an achromatic half-wave plate (HWP) before the focusing doublet. The light transmitted by the sample is collected by photodiode P$_t$ and referenced to the incident light intensity measured by photodiode P$_r$. The normalized transmission spectrum is then obtained by scanning the laser wavelength with and without sample.

\subsection{Experimental results}\label{sec:exp_results}

\begin{figure*}
\includegraphics[width=\columnwidth]{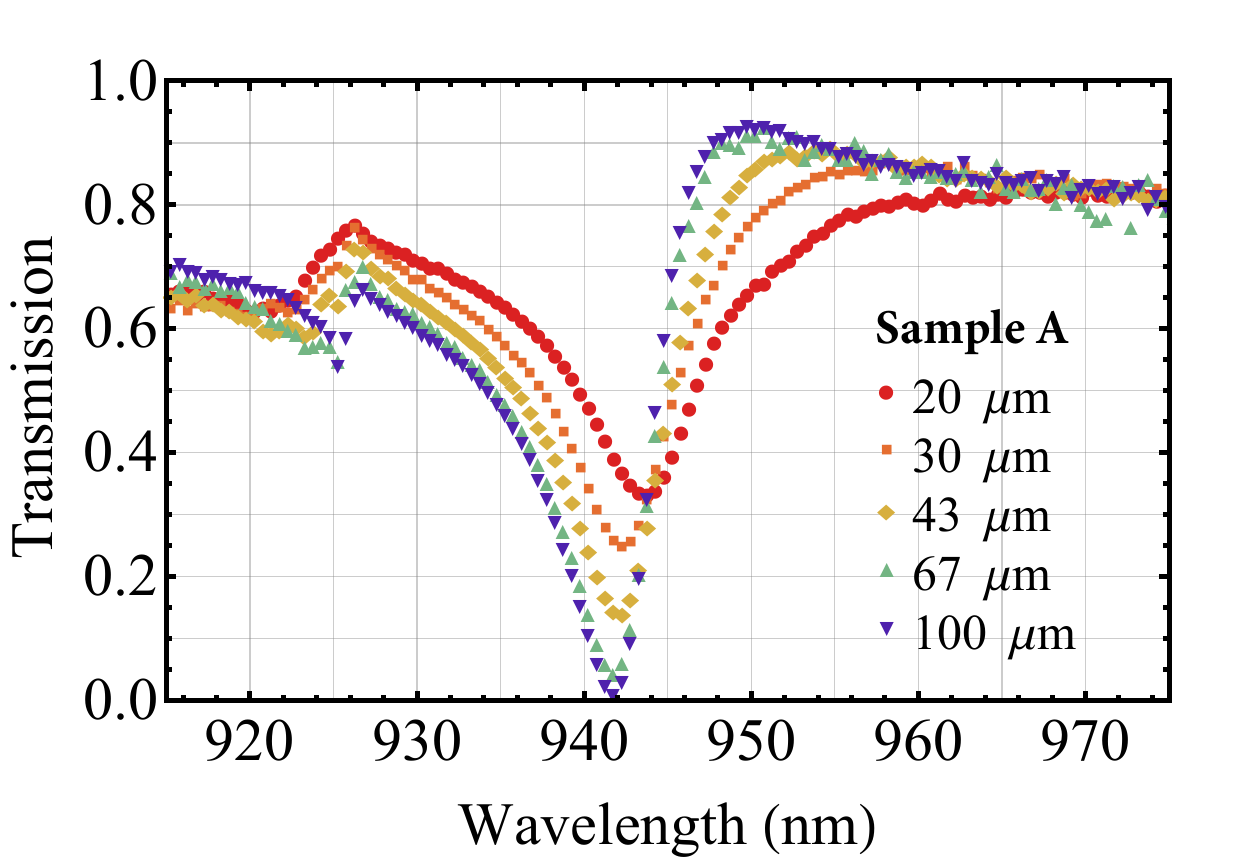}
\includegraphics[width=\columnwidth]{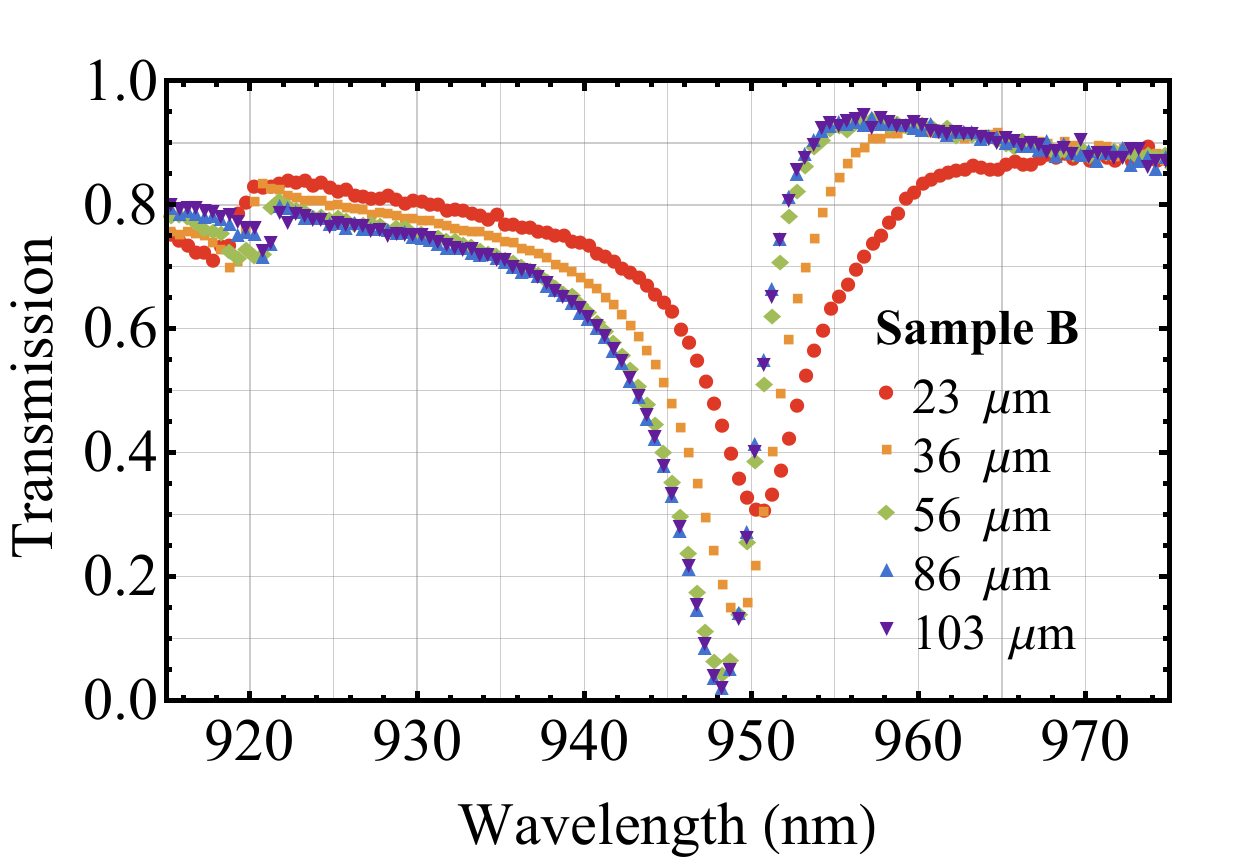}\\
\includegraphics[width=\columnwidth]{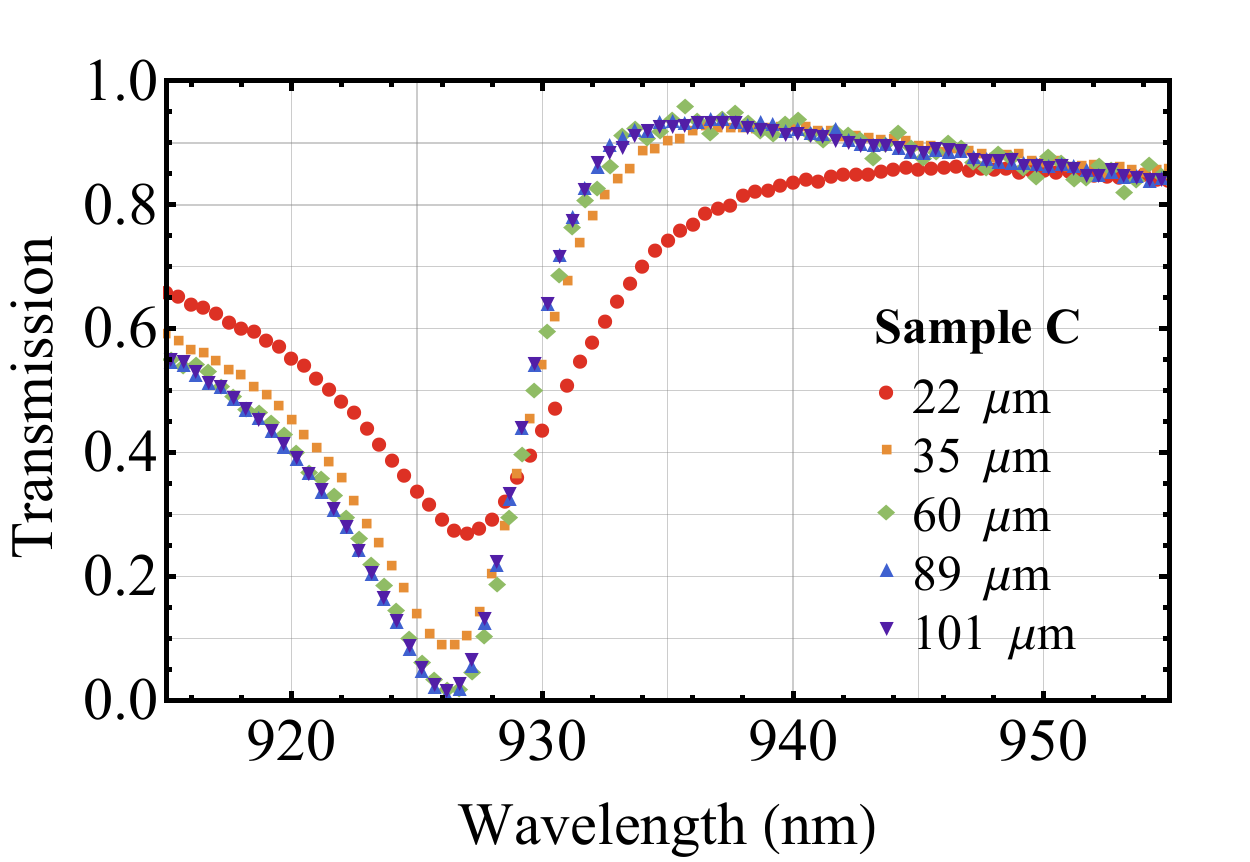}
\includegraphics[width=\columnwidth]{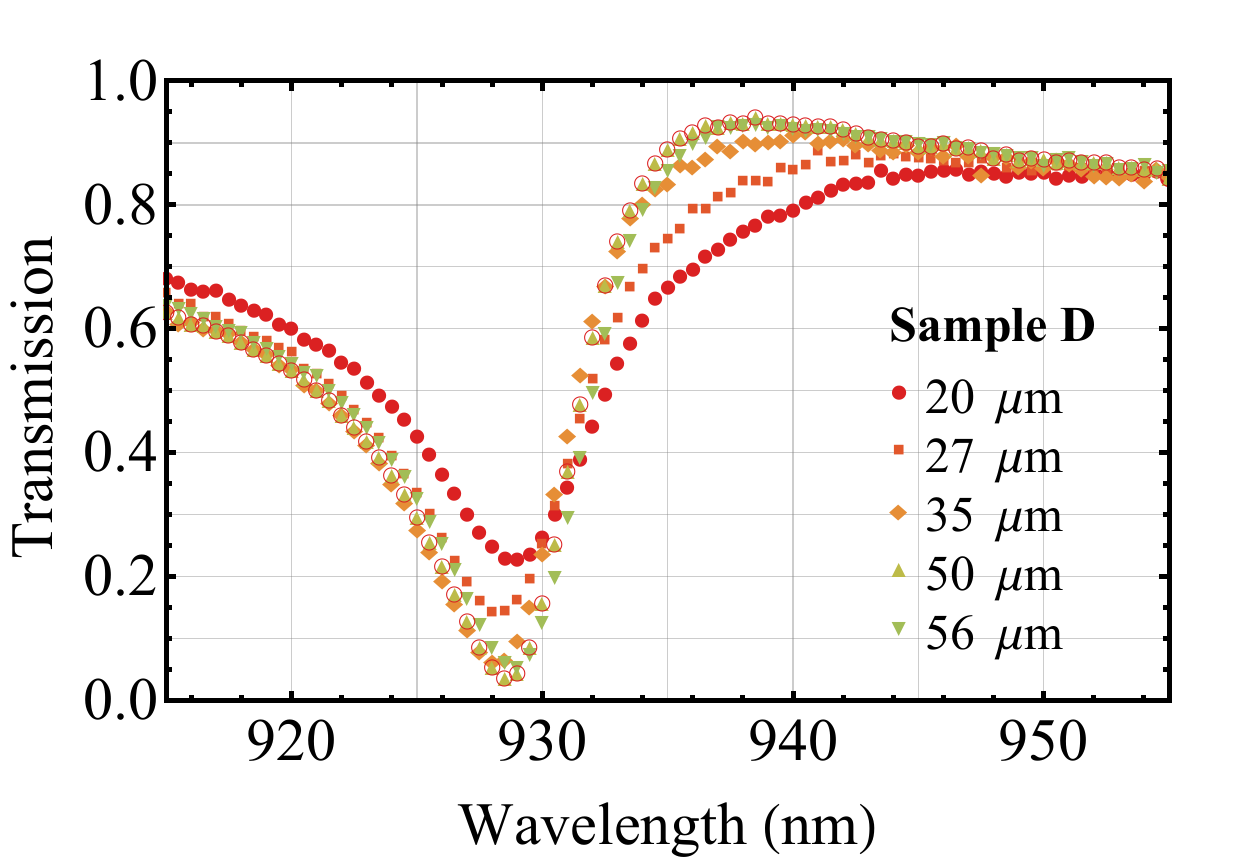}\\
\includegraphics[width=\columnwidth]{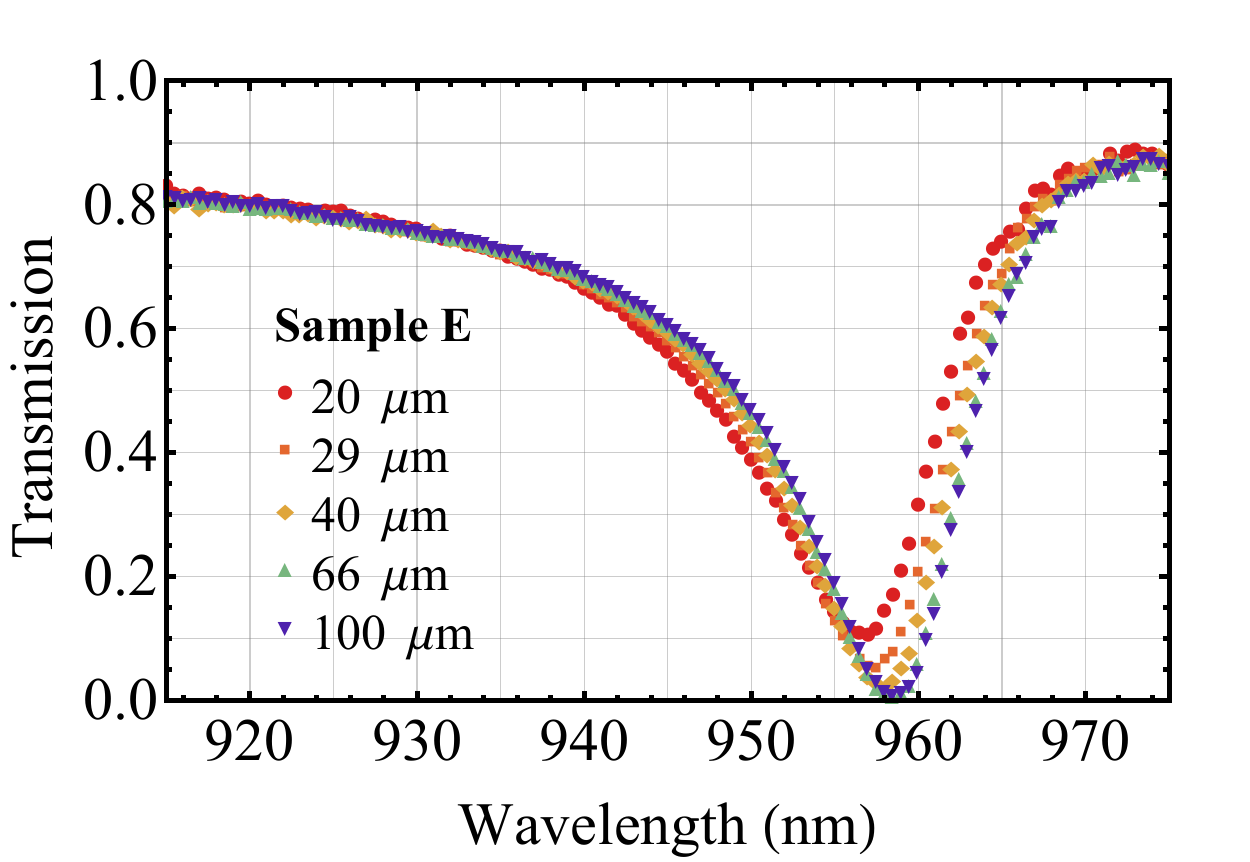}
\caption{Measured normalized transmission spectra of samples A-E for different incoming beam waists $w_0$.}
\label{fig:exp}
\end{figure*}

Figure~\ref{fig:exp} shows the measured normalized transmission spectra of the five gratings whose parameters are given in Table~\ref{tab:tab} for different waists of the incoming beam. All spectra exhibit Fano resonances showing a significant drop of the otherwise high ($\sim 90$ \%) transmission in this wavelength range. Shifts and broadening of these resonances are also clearly observed when the incoming beam is reduced. Concomittantly, the minimum transmission level increases. A second, smaller Fano resonance is observed at a lower wavelength than the main resonance for samples A and B. The main resonance corresponds to the resonant interaction of the incoming light with the guided mode possessing an even symmetry with respect to the grating symmetry plane, while the second resonance corresponds to excitation of the odd symmetry mode. For a perfect infinite grating illuminated by a plane wave the odd symmetry mode cannot be excited at normal incidence. However, for a finite grating illuminated by a Gaussian beam, coupling with this mode becomes possible. We examine the role of collimation and finite size effects on these resonances in detail in the next section. The second resonance is out of the available wavelength range of the laser for samples C, D and E, but its position can be reliably estimated using RCWA simulations, as will be discussed in the next section. For samples C and D the second Fano resonance also occurs at a lower wavelength than the main resonance. The coupling into this mode, increased by collimation and finite-size effects, results in a shift of the main resonance towards higher wavelengths, as observed. For sample E, the second resonance occurs at a higher wavelength then the main resonance, and consequently, the main resonance is shifted towards lower wavelengths, as the beam waist is decreased.

Based on the measured spectra and anticipating on the forthcoming analysis, one can compare the sensitivity of the different samples with respect to collimation and finite-size effects. For instance, the main difference between samples A and B is the finger depth (113 $\mu$m vs 153 $\mu$m). Sample B's deeper fingers result in a narrower Fano resonance, which suggests a lower coupling with the guided mode, and subsequently, a higher sensitivity to finite-size effects. However, the higher wavelength separation between the odd- and even-mode symmetry resonances suggests comparatively smaller collimation effects for sample B than for sample A. As a result, the experimental data show similar performances of both gratings for comparable waists, with a minimum transmission level down to 1.2 \% achieved for the largest waists.

The main difference between samples C and D is the size of the SWG (200 $\mu$m vs 100 $\mu$m), which limits the size of the beam that can be focused onto the grating before diffraction effects alter the transmission. As such, the minimum transmission level of sample D is limited to 3.5\% for a waist of 50 $\mu$m (the miminum transmission level actually increases when the waist is increased from 50 to 56 $\mu$m, as the beam starts spreading over the nonpatterned area). In contrast, the minimum resonant transmission level of sample C decreases down to 1.5\% for a 101 $\mu$m waist. 

The main difference between sample E and the other samples is the much shorter period (646 nm vs 800 nm) required to observe a resonance for TE-polarized light in the same wavelength range. The  broader Fano resonance and the more localized interaction with the guided mode results in a comparatively lower sensitivity to both finite-size and collimation effects, as compared to the TM-polarized light resonant gratings. The minimum transmission levels observed for sample E are thus markedly lower (about 10\% for a 20 $\mu$m waist beam and down to 0.8\% at large waists), as expected.

A quantitative analyse of the respective magnitude of the effects at play in the various structures is given further in Sec.~\ref{sec:analysis} based on the model which will be introduced in the next section.

\section{Theoretical model}\label{sec:model}

\subsection{Plane-wave decomposition}

\begin{figure}[h]
\includegraphics[width=\columnwidth]{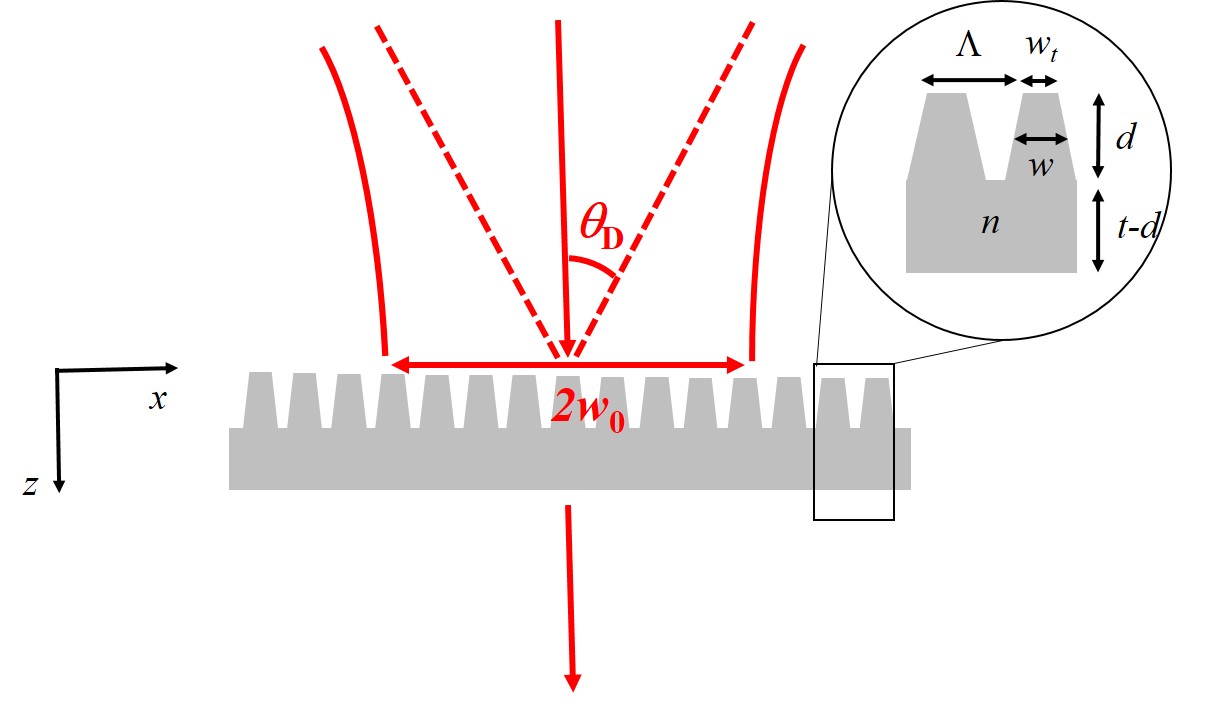}
\caption{Suspended SWG geometry considered: a Gaussian beam (waist $w_0$, divergence angle $\theta_D=\lambda/\pi w_0$) impinges at normal incidence on a suspended subwavelength grating consisting of periodic trapezoidal fingers (pitch $\Lambda$, finger depth $d$, top and mean finger width $w_t$ and $w$) on a slab waveguide with thickness $t-d$. The incident light is linearly polarized either in the $x$-direction (TE) or the $y$-direction (TM).}
\label{fig:SWG}
\end{figure}

We consider a two-layer grating as shown in Fig.~\ref{fig:SWG}, consisting of a trapezoidal one-dimensional grating, infinite in the $y$-direction and periodic in the $x$-direction with period $\Lambda$, finger depth $d$ and top and mean finger widths $w_t$ and $w$, respectively. The grating is supported by a waveguide slab with thickness $t-d$. The refractive index of both the grating fingers and underlying slab is $n$. The structure is surrounded by air ($n=1$) and illuminated by monochromatic light (wavelength $\lambda=2\pi/k$). We assume the beam to be linearly polarized with either TE or TM polarization and impinging at normal incidence on the grating. Within the paraxial approximation the beam is assumed to be Gaussian and weakly focused on the grating with a waist $w_0$ at the grating position smaller than the size of the grating. Two effects due to the nature of the focused beam then affect the transmission of the beam by the SWG in the vicinity of a guided mode resonance: \textit{collimation effects} due to the angular divergence of the beam and \textit{finite-size effects} due to the finite spatial overlap between the incoming mode and the guided modes in the SWG.

In this one-dimensional situation we make use a plane-wave decomposition of the incident field and introduce the Fourier transform of the field amplitude
\begin{equation}
\tilde{E}_\textrm{in}(k_x)=\int_{-\infty}^{\infty} dx\, E_\textrm{in}(x)e^{-i k_xx},
\end{equation}
where $x$ is the spatial coordinate and $k_x=k\sin\theta$ the spatial frequency coordinate. The angular spectrum of the transmitted field is obtained by multiplying the angular spectrum of the incident field by the transfer function of the SWG, $t(k_x)$, and its inverse Fourier transform yields the the spatial transmitted field amplitude
\begin{equation}
E_\textrm{tr}(x)=\frac{1}{2\pi}\int_{-\infty}^{\infty}dk_x\,t(k_x)\tilde{E}_\textrm{in}(k_x)e^{ik_xx}.
\label{eq:E_tr}
\end{equation}

For a fundamental Gaussian beam amplitude
\begin{equation}
E_\textrm{in}(x)=E_0e^{-x^2/w_0^2},
\end{equation}
one has
\begin{equation}
\tilde{E}_\textrm{in}(k_x)=E_0\sqrt{\pi}w_0e^{-(k_xw_0/2)^2}.
\label{eq:E_in_kx}
\end{equation}

The normalized transmission spectrum is then obtained by the ratio of the integral--in real or frequency space--of the squared modulus of the transmitted field amplitude to that of the incident field amplitude, e.g.
\begin{equation}
\mathcal{T}=\frac{\int_{-\infty}^{\infty}dk_x|t(k_x)\tilde{E}_\textrm{in}(k_x)|^2}{\int_{-\infty}^{\infty}dk_x|\tilde{E}_\textrm{in}(k_x)|^2}.
\label{eq:Tav}
\end{equation}

\subsection{Collimation effects}

The transmission function of a plane wave with incidence angle $\theta$ for a weakly focused beam on an infinite, lossless grating can be derived within the frame of the coupled-mode model of Bykov et al.~\cite{Bykov2015} and can be expressed as
\begin{equation}
t_\textrm{CM}(k_x)=t_d\frac{(k-k_0)(k-k_2)-\nu k_x^2}{(k-k_1-i\delta)(k-k_2)-\nu k_x^2},
\label{eq:t_CM}
\end{equation}
where $t_d$ is the off-resonant transmission coefficient, $k_0$, $k_1$ and $k_2$ correspond respectively to the minimum transmission wavenumber and the even and odd guided mode resonant wavenumbers at normal incidence. $\delta$ defines the width of the guided mode and $\nu$ is related to the guided mode group velocity. At normal incidence ($k_x=0$), only the even symmetry guided mode can be excited and the transmission function yields a single-resonance Fano profile with zero minimum transmission at $k=k_0$, a width determined by $\delta$ and a background level determined by $t_d$. At oblique resonance, the odd-symmetry mode can also be excited and the transmission profile displays two Fano resonances, around $k_1$ and $k_2$ for small incidence angles.

The magnitude of collimation effects on the transmission spectrum can thus be evaluated by performing the average stipulated by Eq.~(\ref{eq:Tav}) using the transmission given by Eq. (\ref{eq:t_CM})--or any equivalent simulated or experimentally determined spectra of oblique incident plane waves, as we will show in the next section. As can be seen from Eq.~(\ref{eq:t_CM}), the sensitivity of the spectrum to collimation effects depends on both the parameter $\nu$ and the  separation between the guided mode resonance wavenumbers $k_1$ and $k_2$. For small incidence angles and narrow resonances, such as those considered here, the minimum transmission resonance shift of a plane wave with incidence angle $\theta$ with respect to a plane wave at normal incidence and resonant at $k_0$ is approximately given by $\nu_\theta\theta^2$ with
\begin{equation}
\nu_\theta=\nu\frac{k_0}{k_2-k_0}.
\end{equation}
This quantity will be used as a figure of merit to discuss the sensitivity of a given grating to collimation effects in Sec.~\ref{sec:analysis}.

\subsection{Finite-size effects}

\begin{figure}
\includegraphics[width=0.5\columnwidth]{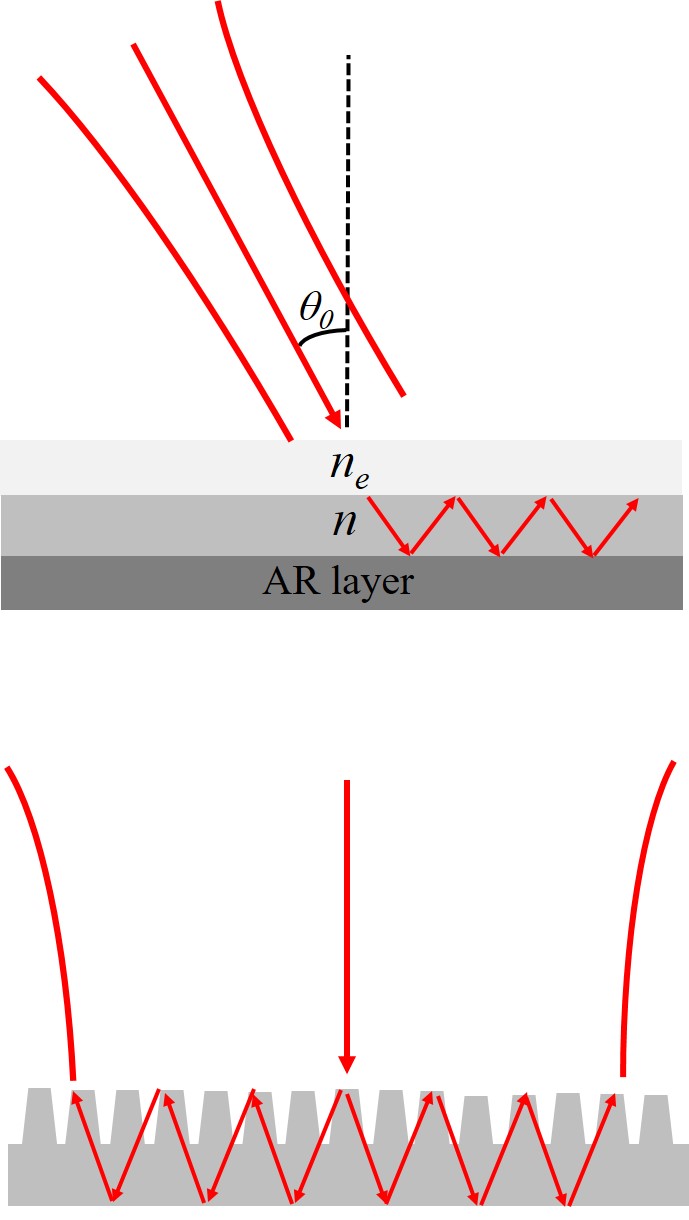}
\caption{Top: AR-coated waveguide illuminated at oblique incidence. Bottom: suspended SWG illuminated at normal incidence.}
\label{fig:AR}
\end{figure}

To estimate the magnitude of the finite size effects, we follow the approach of Jacob, Duun and Moharam~\cite{Jacob2000,Jacob2001}, which exploits the effective medium representation of the grating layer on top of a waveguide slab to treat the interference between the incident and guided modes. Under illumination at oblique incidence and using an additional AR-coating layer under the waveguide slab (Fig.~\ref{fig:AR}), this model allows for a simple analytical determination for the guided mode resonances and of the angular and spectral dephasing rates of the guided modes, and thereby allows for a quantitative discussion of finite-size effects.

\subsubsection{Oblique incidence}
For an infinite grating illuminated at oblique incidence and for a suitable AR-layer design, the reflection response of the structure can be expressed as
\begin{equation}
r_\textrm{wav}^{(\infty)}=\eta e^{i\phi}\sum_{n=0}^{\infty}(1-\eta)^ne^{in\phi}=\frac{\eta e^{i\phi}}{1-(1-\eta)e^{i\phi}},
\label{eq:r_w}
\end{equation}
where $\eta$ is the effective grating coupling strength, equal to the diffraction efficiency of the first order in the waveguide region, and where $\phi$ is the round-trip phase of the diffracted order as it propagates in the structure. For a guided mode to exist in the waveguide slab, the total phase shift experienced by the wave traveling from the upper film interface to the lower film interface and back again must be an integer multiple of $2\pi$, which is achieved when the tangential component of the diffracted order matches the propagation constant of a mode supported by the structure
\begin{equation}
\beta=k\sin\theta_0\pm\frac{2\pi}{\Lambda}.
\end{equation}
The characteristic equation for the slab waveguide is then
\begin{equation}
\phi=2\kappa (t-d)+2\phi_c+2\phi_s=2m\pi,
\label{eq:res}
\end{equation}
where $m$ is the diffracted mode number, 
\begin{equation}
\kappa=\sqrt{n^2 k^2-\beta^2}
\label{eq:kappa}
\end{equation} is the transverse propagation constant of the waveguide. The phase shifts associated with reflection from the film/air interface and from the film/grating interface are given by
\begin{align}
\phi_s&=\tan^{-1}\left(n^{2\rho}\frac{\gamma}{\kappa}\right),\label{eq:phi_s}\\
\phi_c&=\tan^{-1}\left[n^{2\rho}\frac{\gamma_e}{\kappa}\tanh\left[\tanh^{-1}\left(\frac{\gamma}{\gamma_e}\right)+\gamma_ed\right]\right],\label{eq:phi_c}
\end{align}
with
\begin{align}
\gamma&=\sqrt{\beta^2-k^2},\\ 
\gamma_e&=\sqrt{\beta^2-n_e^2k^2},
\end{align} and where 
\begin{equation}
\rho=0\,\, \textrm{and}\,\, n_e=\sqrt{n^2f+1-f}
\label{eq:nTE}
\end{equation} for TE-polarized light, and
\begin{equation}
\rho=1\,\,\textrm{and}\,\, n_e=1/\sqrt{f/n^2+1-f}
\label{eq:nTM}\end{equation} for TM-polarized light.

For a low effective grating coupling strength the angular response (\ref{eq:r_w}) around a resonance predicted by Eq.~(\ref{eq:res}) is approximately Lorentzian with an angular linewidth (FWHM) $\Delta\phi\simeq 2\eta$. From Eqs.~(\ref{eq:res},\ref{eq:kappa},\ref{eq:phi_s},\ref{eq:phi_c}) the spectral and modal dephasing rates, $d\phi/d\lambda$ and $d\phi/d\beta$, can also be evaluated close to resonance. The angular linewidth is then directly related to the spectral linewidth by
\begin{equation}
\eta=\delta\frac{\lambda_0^2}{2\pi}\left|\frac{d\phi}{d\lambda}\right|_{k=k_0}.
\label{eq:eta}
\end{equation}

For a finite-length structure or an interaction with a finite-size beam there is a finite number $N$ of waveguide reflections in the region of overlap between the incident beam and the guided mode, such that the infinite summation leading to (\ref{eq:r_w}) must be truncated after $N$ reflections. The SWG reflectivity coefficient can thus be written as
\begin{align*}
r_\textrm{wav}^{(N)}&=\eta e^{i\phi}\sum_{n=0}^{N}(1-\eta)^ne^{in\phi}\\
&=\frac{\eta e^{i\phi}}{1-(1-\eta)e^{i\phi}}\left[1-(1-\eta)^Ne^{i N\phi}\right]\\
&=(1-\alpha)r_\textrm{wav}^{(\infty)},
\end{align*}
i.e. the product of the infinite structure/beam coefficient with a correction factor $1-\alpha$, where 
\begin{equation}
\alpha=(1-\eta)^Ne^{iN\phi}.
\label{eq:alpha}
\end{equation} 
The number of diffracted components $N=L/l$ is given by the ratio of the interaction length $L$ to the round-trip length $l$ of the diffracted order in the waveguide, which is equal to the modal dephasing at resonance
\begin{equation}
l=\left|\frac{d\phi}{d\beta}\right|_{k=k_0}.
\label{eq:l}
\end{equation}

\subsubsection{Coupled-mode model including collimation and finite-beam effects}

The situation investigated in the experiments differs from the previous finite waveguide model in mainly two aspects: first, no AR-layer is used, and second, the grating is illuminated at normal incidence, as depicted in Fig.~\ref{fig:AR}.

In absence of an AR-coating layer, the off-resonant reflection/transmission levels at normal incidence are in general non-zero/non-unity. A generic expression for the transmission of a plane wave resulting from the interference between the guided mode (characterized by the resonant wavenumber $k_1$ and resonance width $\delta$) and the direct transmission through the slab (characterized by the transmission coefficient $t_d$) is given by~\cite{Fan2002,Fan2003}
\begin{equation}
t_{\infty}=t_d+\frac{a}{(k-k_1)-i\delta}  \equiv t_d\frac{k-k_0}{k-k_1-i\delta}
\label{eq:t_inf0}
\end{equation}
which is equivalent to the coupled-mode transfer function $t_\textrm{CM}(k_x=0)$ given by Eq.~(\ref{eq:t_CM}) for a normally incident plane-wave.

The interference between the guided mode and a finite-size beam with waist $w_0$ impinging on the grating is then phenomenologically included by multiplying the guided mode term by the correction factor previously introduced in the oblique incidence situation
\begin{align}
t_\textrm{fin}&=(1-\alpha)t_{\infty}+\alpha t_d\nonumber\\
&=t_d+(1-\alpha)\frac{a}{k-k_1+i\delta}\nonumber\\
&=t_d\frac{k-k_0+\alpha\Delta}{k-k_1-i\delta},
\label{eq:t_fin}
\end{align}
where $\Delta=k_0-k_1-i\delta$. 

Taking $L$ to be given by the beam waist $w_0$, one has, close to resonance,
\begin{equation}\alpha\simeq(1-\eta)^{w_0/l}=e^{-w_0/A},\end{equation}
with
\begin{equation}
A=-\frac{l}{\ln(1-\eta)}.
\label{eq:A}
\end{equation}
The exponential variation of the correction factor with the beam waist leads to a reduction of the interference with the guided mode, which results as expected in a shift and broadening of the Fano resonance as well as an increased minimum transmission level. The effect becomes significant when the waist becomes of the order of $A\simeq l/\eta$ for small $\eta$'s. $A$ can thus be used as a figure of merit to discuss the sensitivity of a given grating structure to finite-size effects.

Finally, in order to take into account both collimation and finite beam-size effects the previous models can be phenomenologically combined in order to generalize the coupled-mode model transfer function by including the finite-beam size correction factor in Eq.~(\ref{eq:t_CM})
\begin{equation}
t_\textrm{CM,fin}(k_x)=t_d\frac{(k-k_0+\alpha\Delta)(k-k_2)-\nu k_x^2}{(k-k_1-i\delta)(k-k_2)-\nu k_x^2}.
\label{eq:tCMfin}
\end{equation}
Eq.~(\ref{eq:tCMfin}) will serve as the basis for analyzing the experimental spectra in the following section. It also allows for a straightforward comparison of the magnitude of collimation and finite-size effects, by setting taking the limits $\alpha\rightarrow 0$ and $\nu\rightarrow 0$, respectively.


\section{Analysis}\label{sec:analysis}

\subsection{Analysis for sample A}

\subsubsection{Collimation effects}

\begin{figure}
\includegraphics[width=\columnwidth]{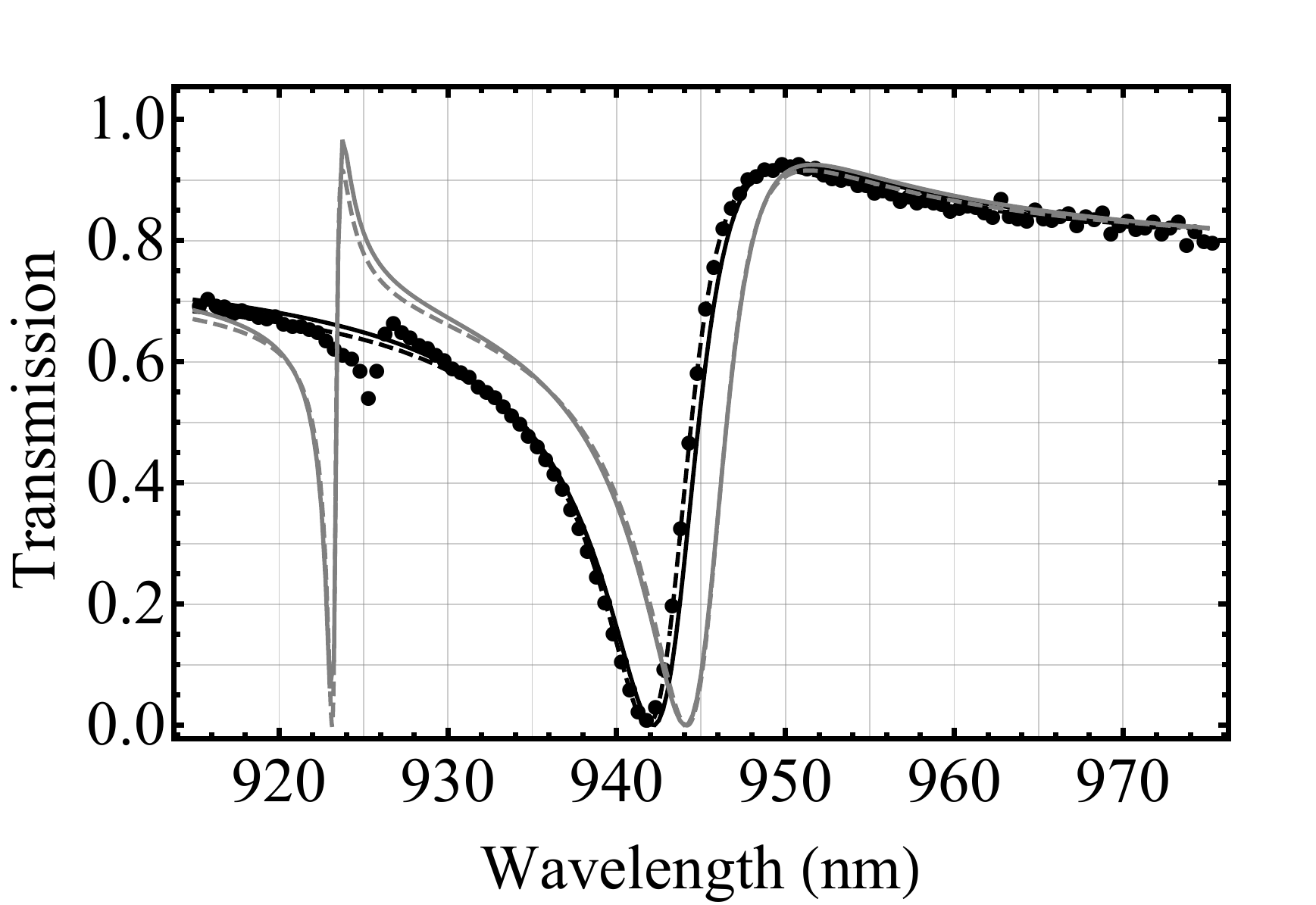}
\caption{Normalized transmission spectra for sample A. Black dots: experimental data for a waist $w_0=100$ $\mu$m. Black plain line: RCWA predictions for normally incident plane wave. Black dashed line: result of a fit of the experimental data with the coupled-mode model for normally incident plane wave. Grey plain line: RCWA predictions for plane wave with incidence angle 0.5$^{\circ}$. Grey dashed line: result of a fit of the experimental data with the coupled-mode model plane wave with incidence angle 0.5$^{\circ}$.}
\label{fig:cal}
\end{figure}

Figure~\ref{fig:cal} shows normalized transmission spectra of sample A illuminated by a normally incident TM-polarized Gaussian beam with a "large" waist of 100 $\mu$m (dots). The measured spectrum is in excellent agreement with the RCWA predictions (dashed) obtained using MIST~\cite{MIST} and based on the independently determined geometrical parameters of the grating and assuming a TM-polarized normally incident plane wave. The plain curve shows the result of a fit of the experimental data with the coupled-mode model predictions for a normally incident plane wave $|t_\textrm{CM}(0)|^2$, yielding $t_d=0.87$, $\lambda_0=941.8$ nm, $\lambda_1=943.2$ nm and $\delta_\lambda=\lambda_0^2/(2\pi)\delta=2.9$ nm.

Figure~\ref{fig:cal} also shows the RCWA predicted spectra of plane waves with incidence angles $\theta=0.5^{\circ}$ (dot-dashed), clearly showing the appearance of the second Fano resonance and the shifts of the resonances with the incidence angles. A fit of such spectra with the coupled-mode model allows for further determining $\lambda_2=925.3$ and $\nu=2.9\times 10^{-5}$ nm$^{-2}$. These values could also be determined experimentally insofar that the patterned area of the SWG is large enough so that one can operate with large enough beam waists and the shifts due collimation/finite-size effects can be neglected (see e.g.~\cite{Parthenopoulos2021} for a study of oblique incidence).

\begin{figure}
\includegraphics[width=\columnwidth]{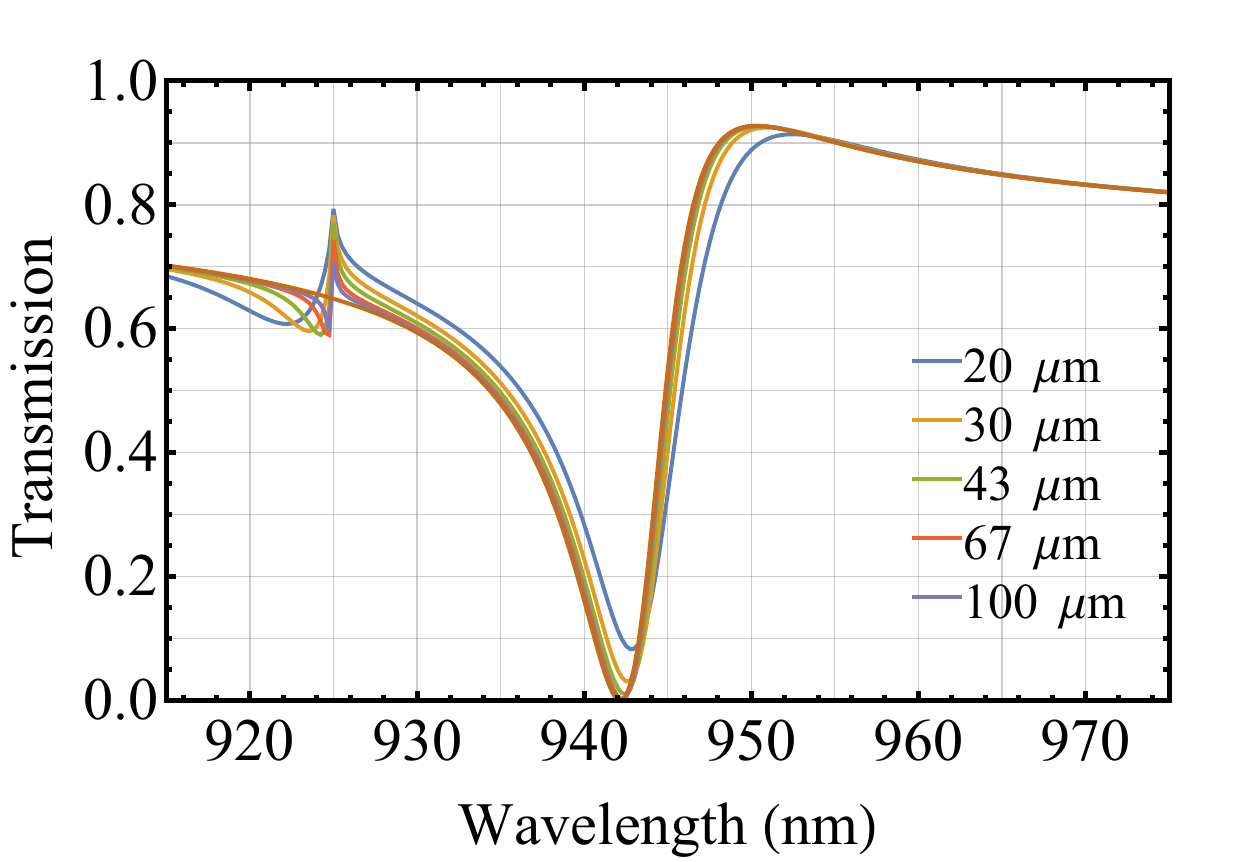}
\includegraphics[width=\columnwidth]{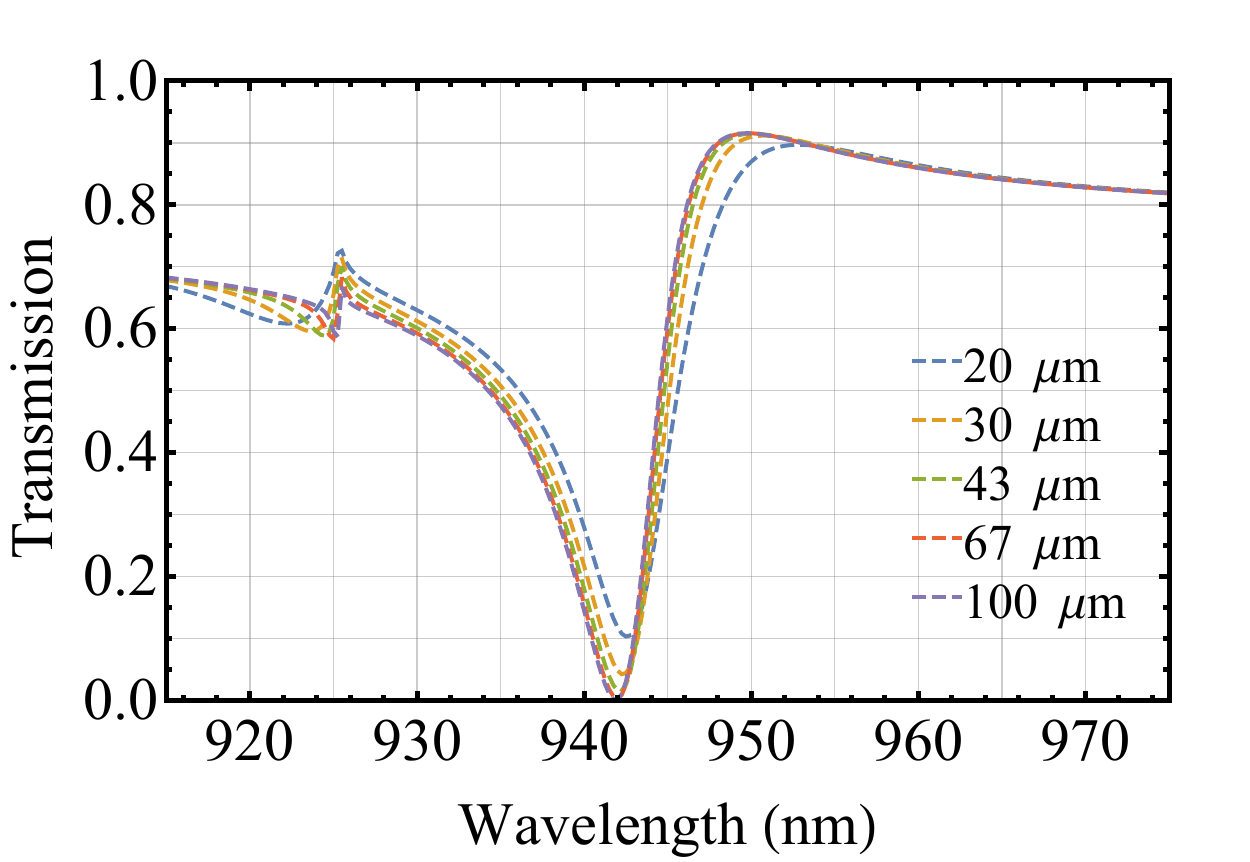}
\caption{\textit{Collimation effects}. Top: RCWA predicted transmission spectra for sample A based on the experimentally determined grating parameters and a Gaussian angular average for different waists $w_0=20$, $30$, $43$, $67$ and $100$ $\mu$m. Bottom: Corresponding coupled-mode model predicted spectra for  $t_d=0.87$, $\lambda_0=941.8$ nm, $\lambda_1=943.2$ nm, $\delta_\lambda=\lambda_0^2/(2\pi)\delta=2.9$ nm, $\lambda_2=925.3$ nm and $\nu=2.9\times 10^{-5}$ nm$^{-2}$.}
\label{fig:mist_bykov}
\end{figure}

To investigate collimation effects for various beam focusings, Fig.~\ref{fig:mist_bykov} compares the transmission spectra predicted by RCWA simulations and the coupled-mode model for different incident beam waists. The spectra are obtained by averaging the RCWA spectra obtained for TM-polarized plane waves impinging on the grating for various angles of incidences according to Eq.~(\ref{eq:E_tr}) with $\tilde{E}_\textrm{in}(k_x)$ given by (\ref{eq:E_in_kx}). The values for the waists used correspond to those used in the experiments. As already pointed out in~\cite{Bykov2015,Parthenopoulos2021}, excellent agreement is observed between the full numerical simulation predictions and those of the coupled mode model. Collimation effects are clearly visible, resulting in a positive shift and broadening of the main Fano resonance, and an increasing mimimum transmission level, as the beam waist is decreased. However, comparing these spectra with the experimentally determined ones (Fig.~\ref{fig:exp}), the magnitude of the predicted collimation effects does not suffice to explain the experimentally observed resonance shifts and broadenings.

\subsubsection{Finite-size effects}

\begin{figure}
\includegraphics[width=\columnwidth]{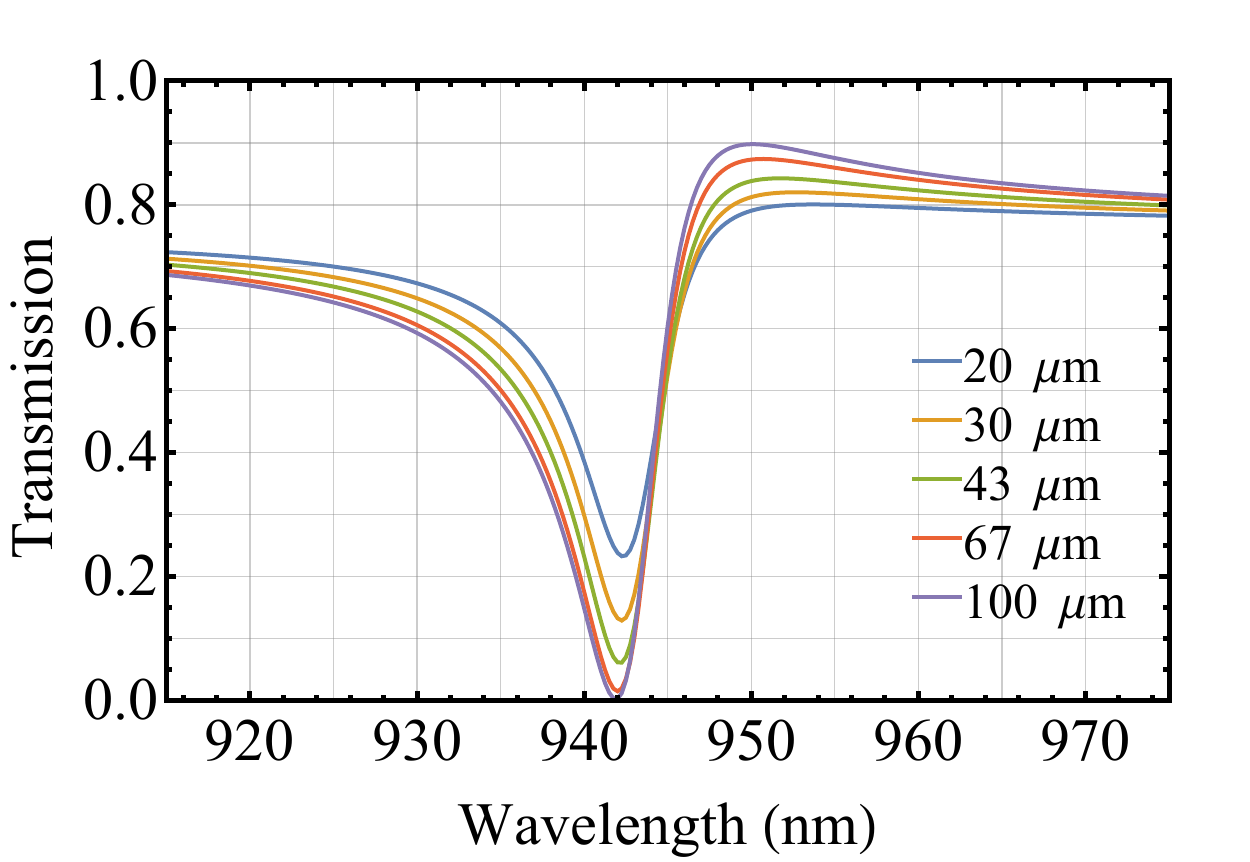}
\caption{\textit{Finite-size effects.} Normalized transmission spectra of sample A predicted by the waveguide model for the same waists as in Fig.~\ref{fig:mist_bykov}.}
\label{fig:fin}
\end{figure}

In order to assess the magnitude of finite-size effects, the parameter $A$ of Eq.~(\ref{eq:A}) must be evaluated. The modal dephasing $l$ can be estimated as follows: first, the value of the grating effective refractive index close to the experimentally determined $k_0$, using Eq.~(\ref{eq:res}) with $k=k_0$, $\theta_0=0$ and $m=0$. For sample A this yields a value of $n_e\simeq 1.31$, somehow larger than the value of 1.21 predicted by Eq.~(\ref{eq:nTM}). An estimate of the value of the modal dephasing is then obtained from Eq.~(\ref{eq:l}) using $k=k_0$ and the value of $n_e$ previously determined, yielding $l\simeq 3.0$ $\mu$m. An estimate of the value of the grating coupling strength is obtained from Eq.~(\ref{eq:eta}) using the experimentally determined $\delta_\lambda$, yielding here $\eta\simeq 8.2$ \%. The transmission spectra predicted by Eq.~(\ref{eq:t_fin}) for differents waists are finally plotted in Fig.~\ref{fig:fin}. The expected resonance shift and broadening and increase in minimum transmission level when the waist is reduced are observed and, for this particular grating, are found to be more significant than those due to collimation effects.

\subsubsection{Comparison between experimental results and coupled-mode model predictions}

\begin{figure}
\includegraphics[width=\columnwidth]{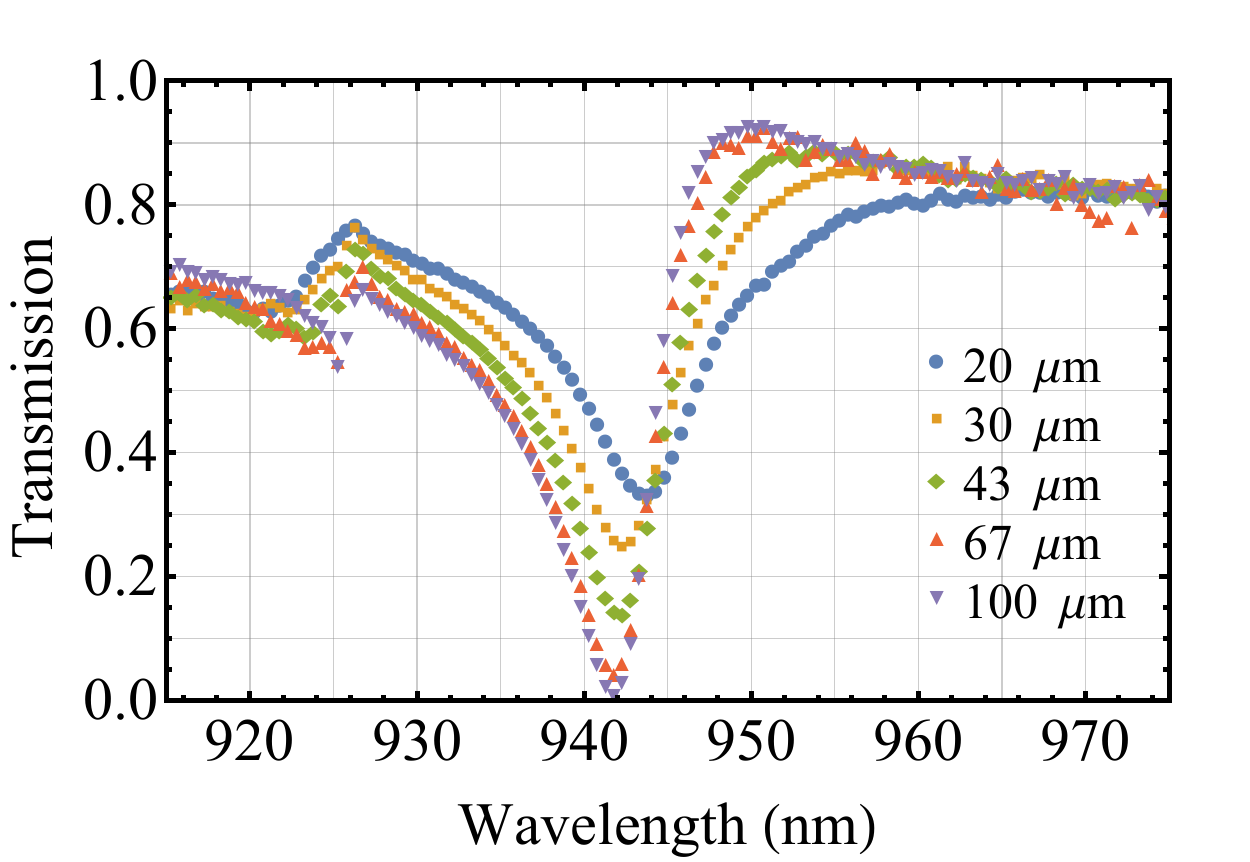}\\
\includegraphics[width=\columnwidth]{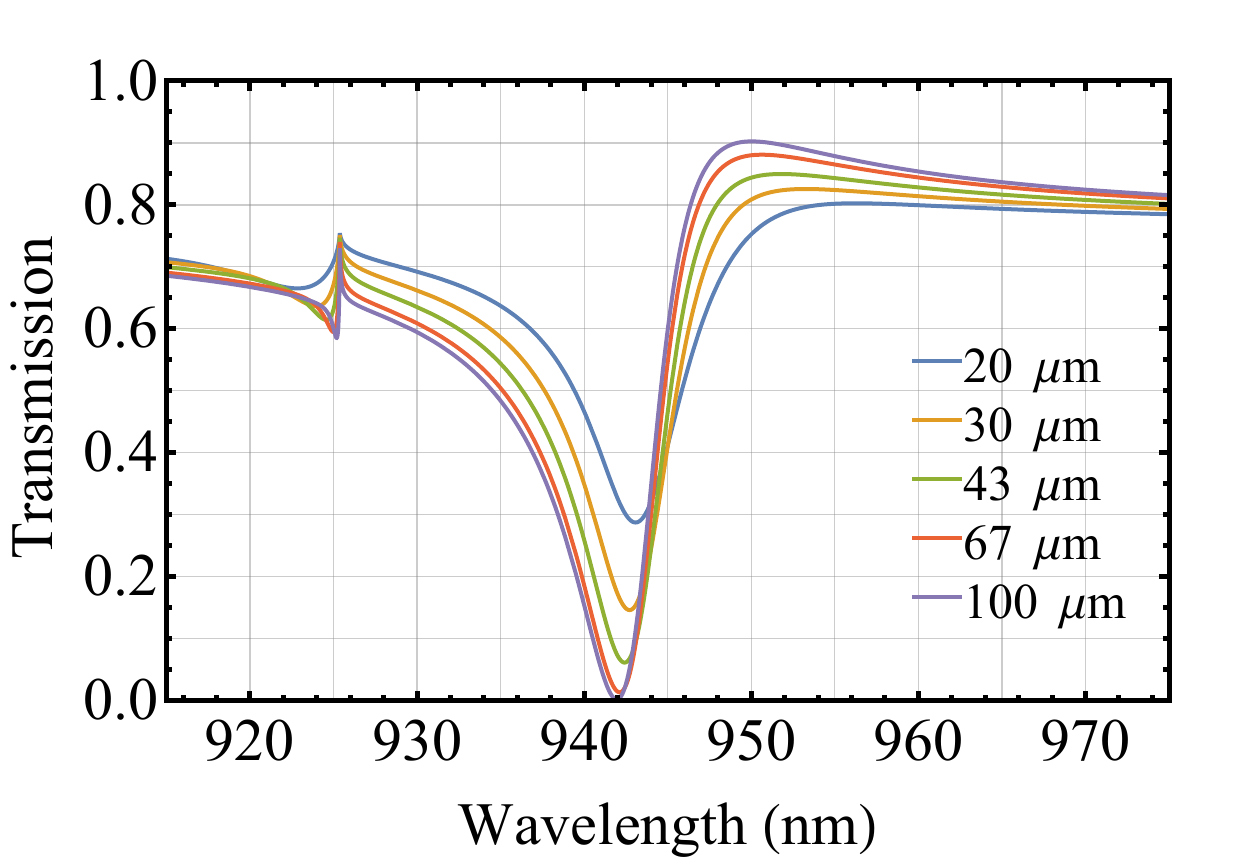}
\caption{Comparison between the experimental transmission spectra of sample A (top) and those resulting from a fit with the phenomenological model including both collimation and finite-size effects (bottom).}
\label{fig:fittot}
\end{figure}

Figure~\ref{fig:fittot} shows a comparison of the experimental spectra of Fig.~\ref{fig:exp}(a) with the results of a global fit of the experimental data with the phenomenological model [Eq.~(\ref{eq:tCMfin})], where $\alpha$ is given by (\ref{eq:alpha}) and $A$ is left as a free parameter, the other parameters having the values determined previously. In addition to the very good agreement between the measured spectrum and the fit result, the resulting fit value, $A_\textrm{fit}=32.0$ $\mu$m, is in reasonable agreement with the estimate based on the approximate waveguide model, $A_\textrm{w}\simeq l/\eta=37.0$ $\mu$m.

\subsection{Analysis for samples B-D}

\begin{figure*}
\includegraphics[width=0.85\columnwidth]{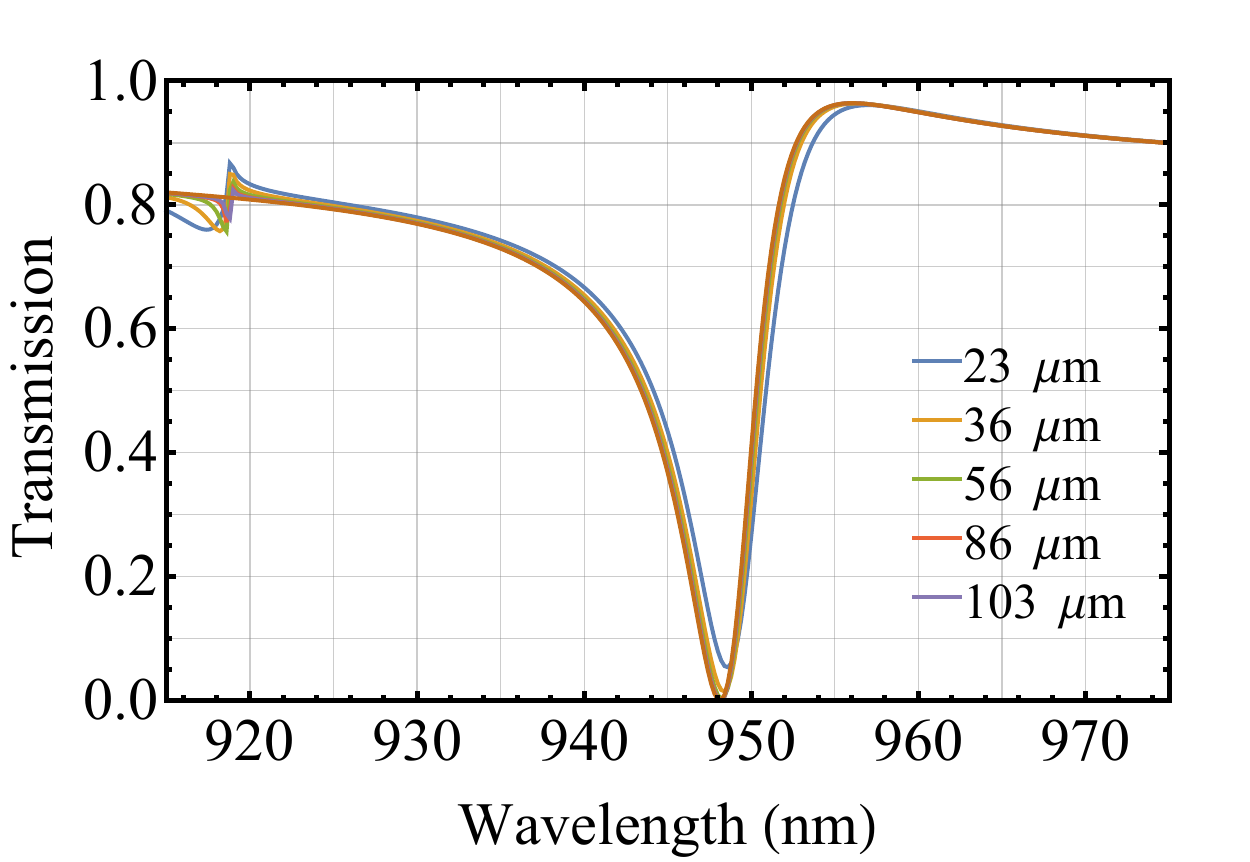}\includegraphics[width=0.85\columnwidth]{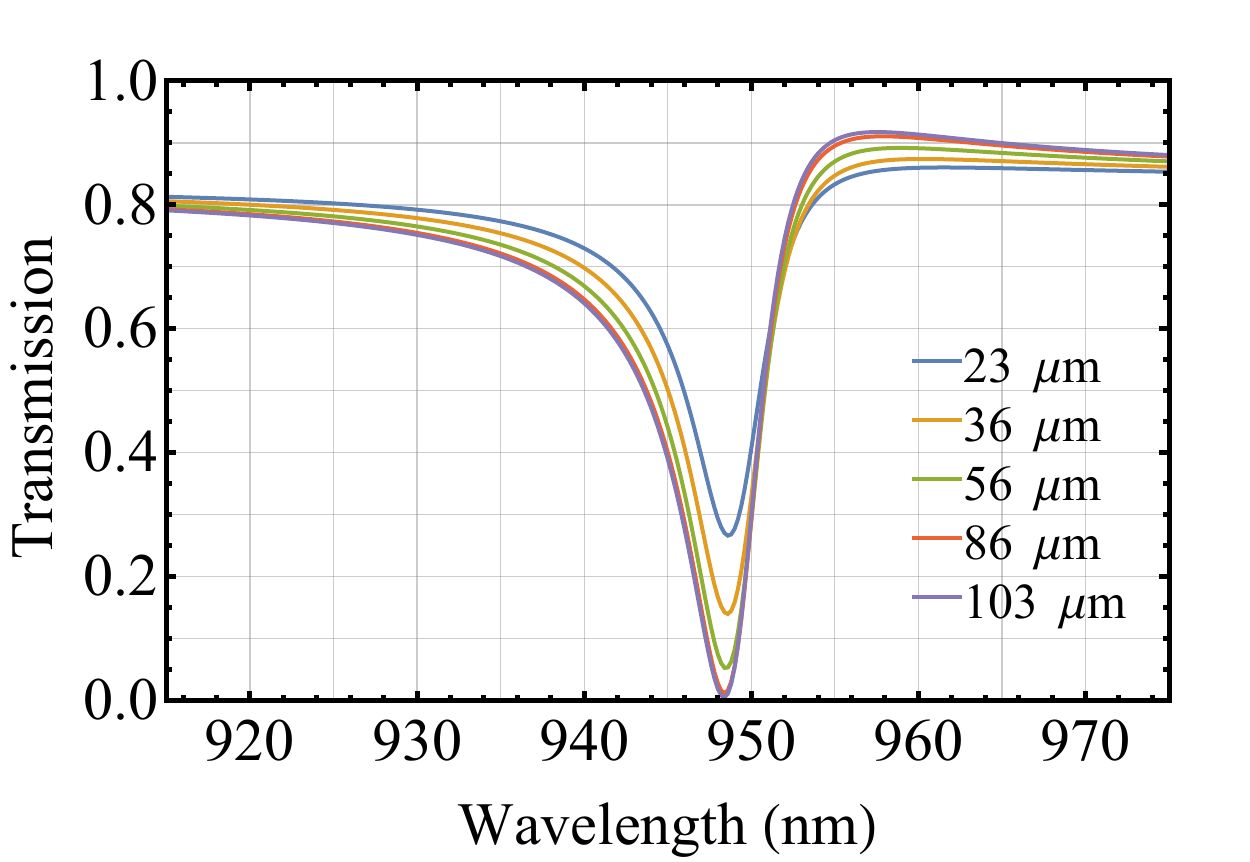}\\
\includegraphics[width=0.85\columnwidth]{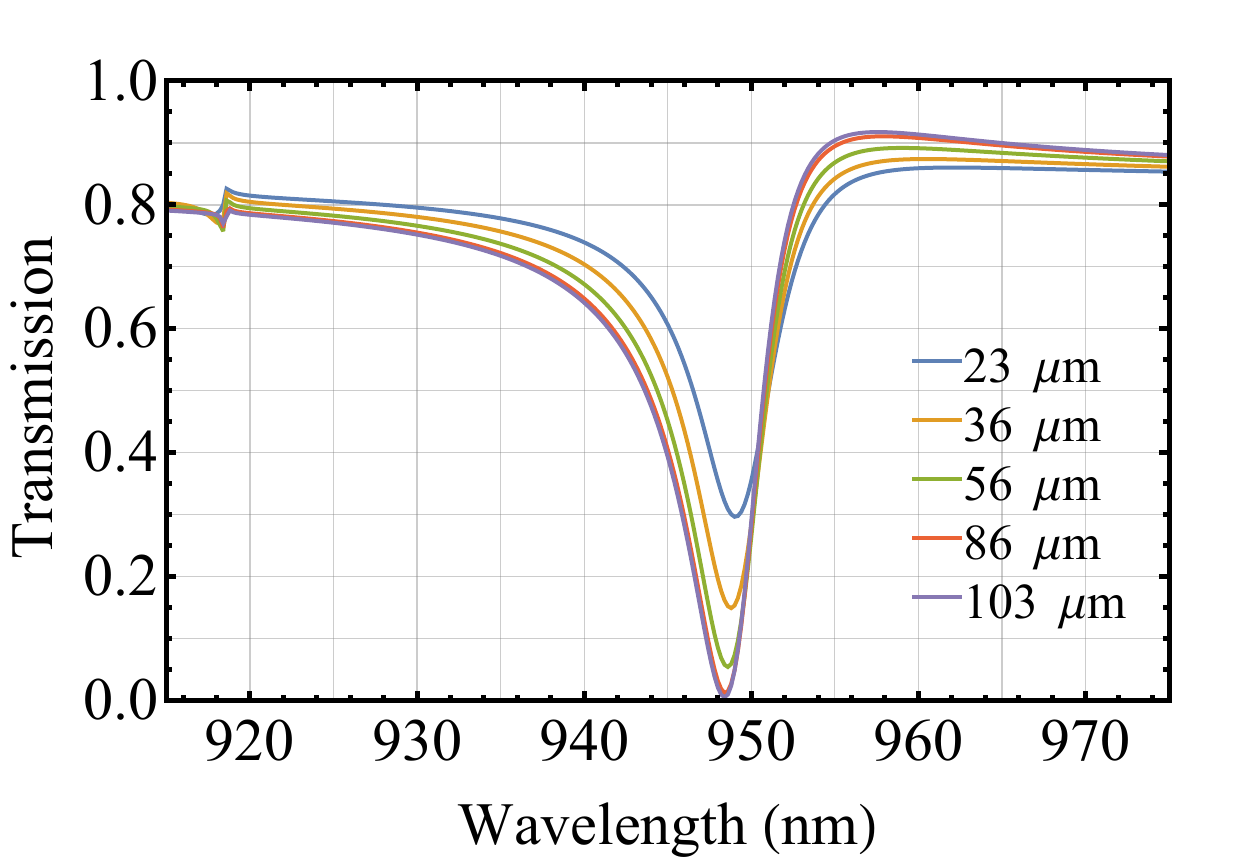}\includegraphics[width=0.85\columnwidth]{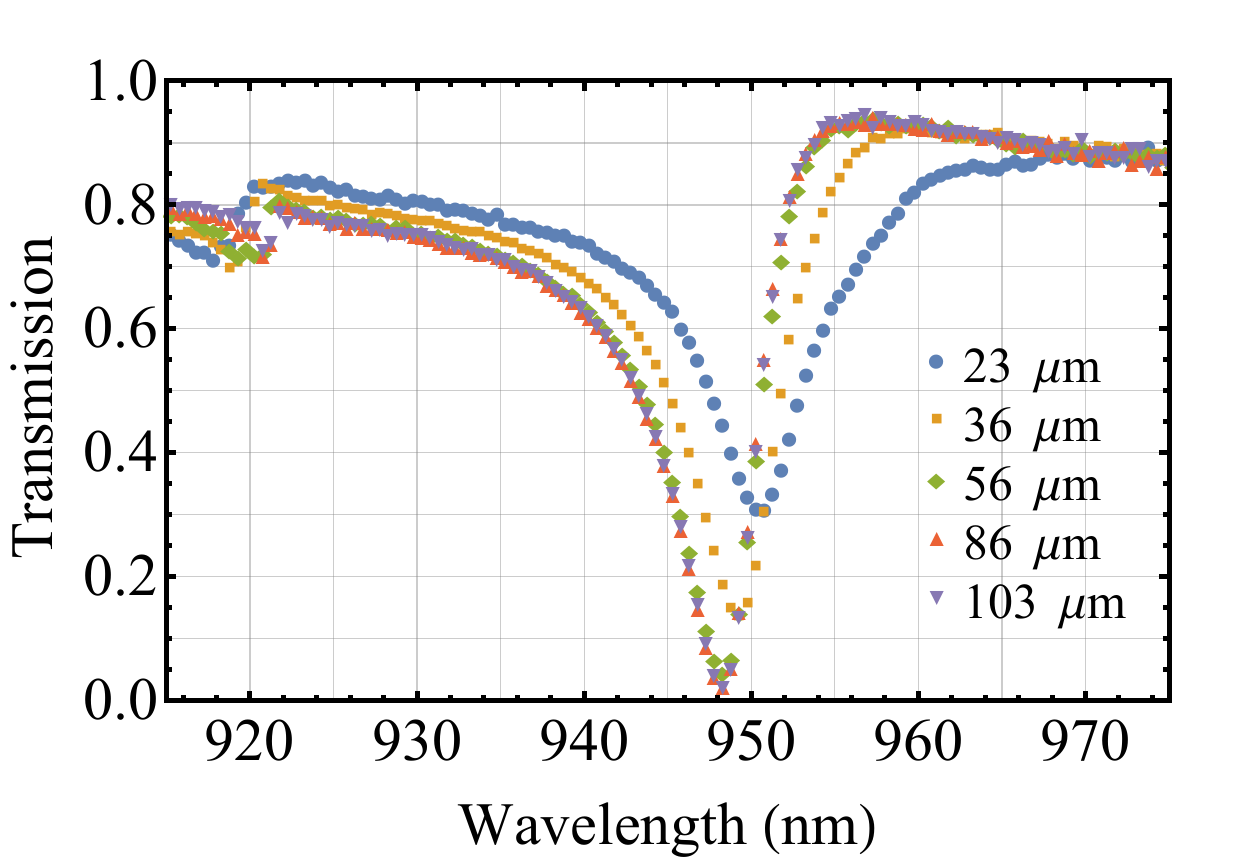}
\caption{Sample B. Simulated transmission spectra including collimation effects only (top left), finite-size effects only (top right) and both effects (bottom left), as well as the corresponding experimental spectra (bottom right).}
\label{fig:B4}
\end{figure*}

\begin{figure*}
\includegraphics[width=0.85\columnwidth]{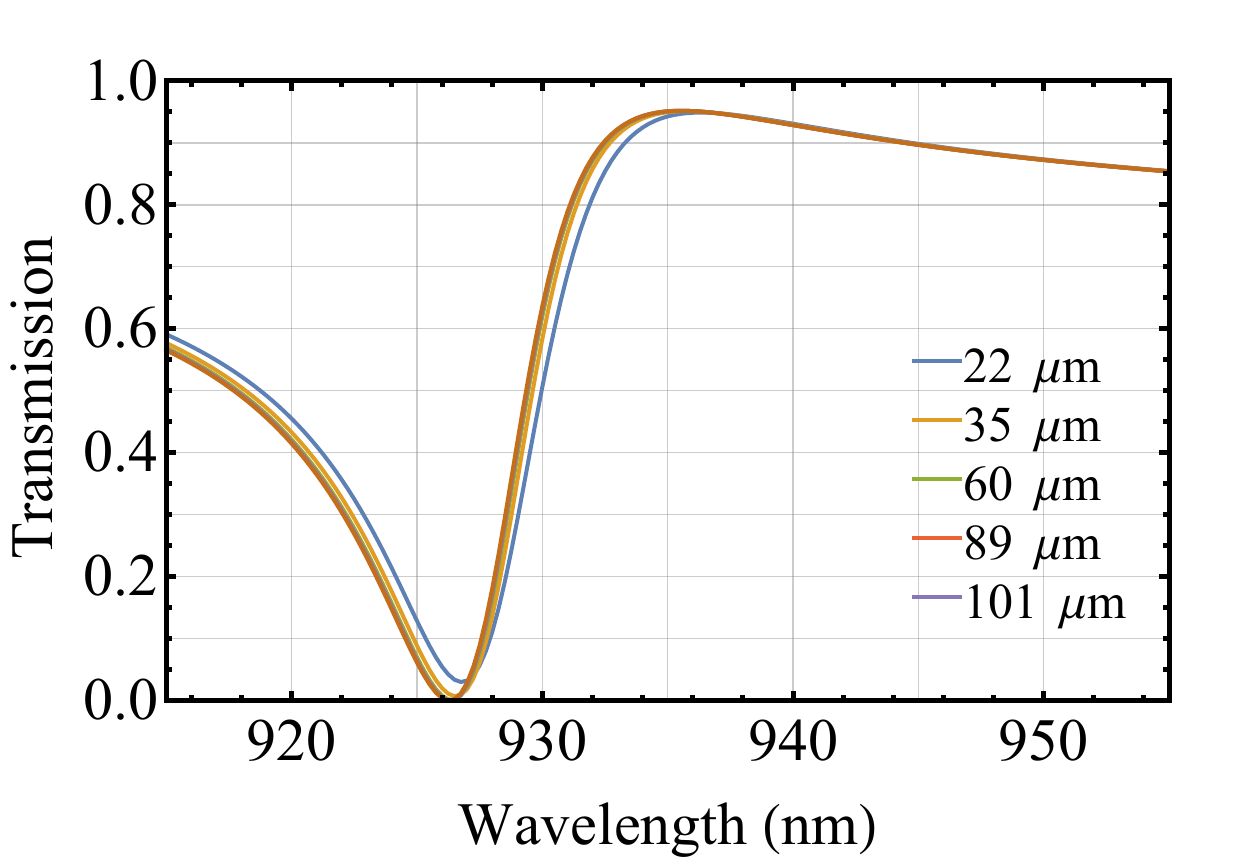}\includegraphics[width=0.85\columnwidth]{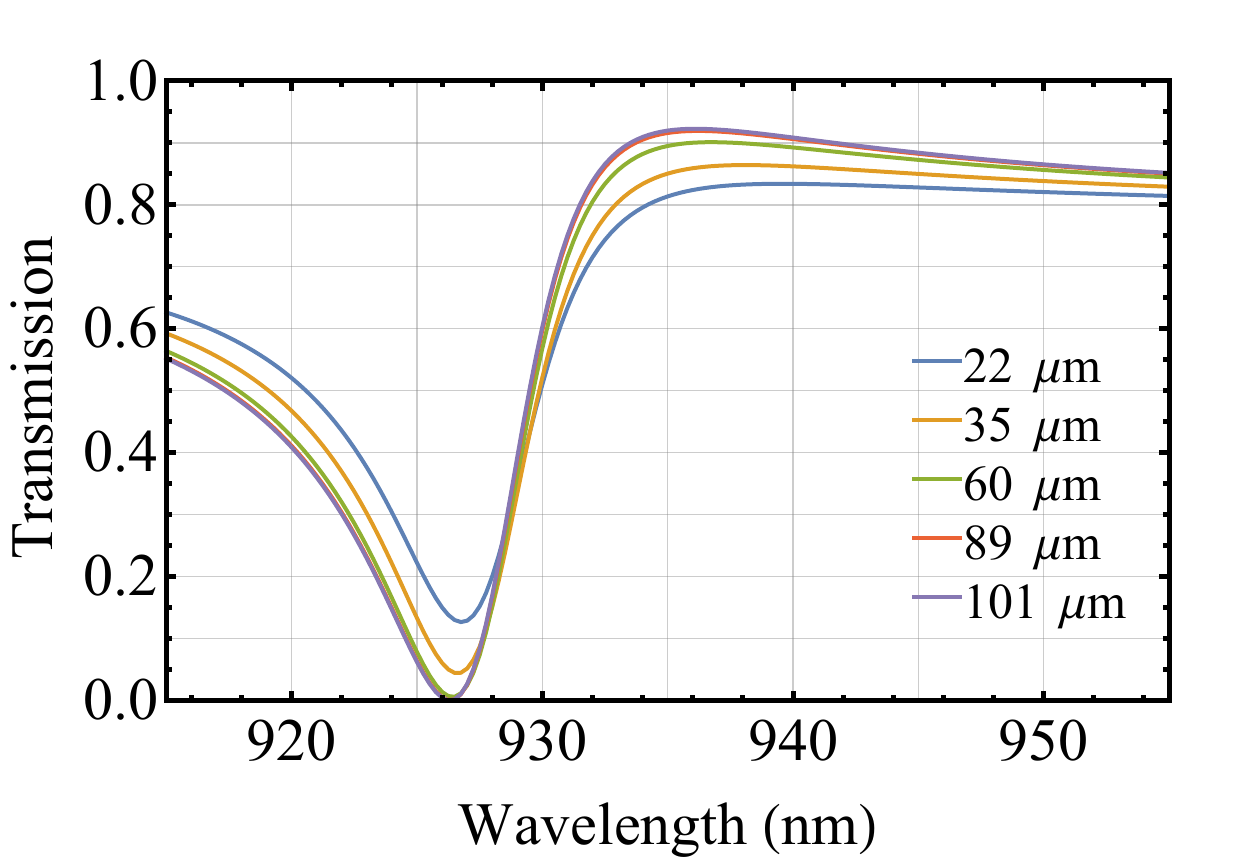}\\
\includegraphics[width=0.85\columnwidth]{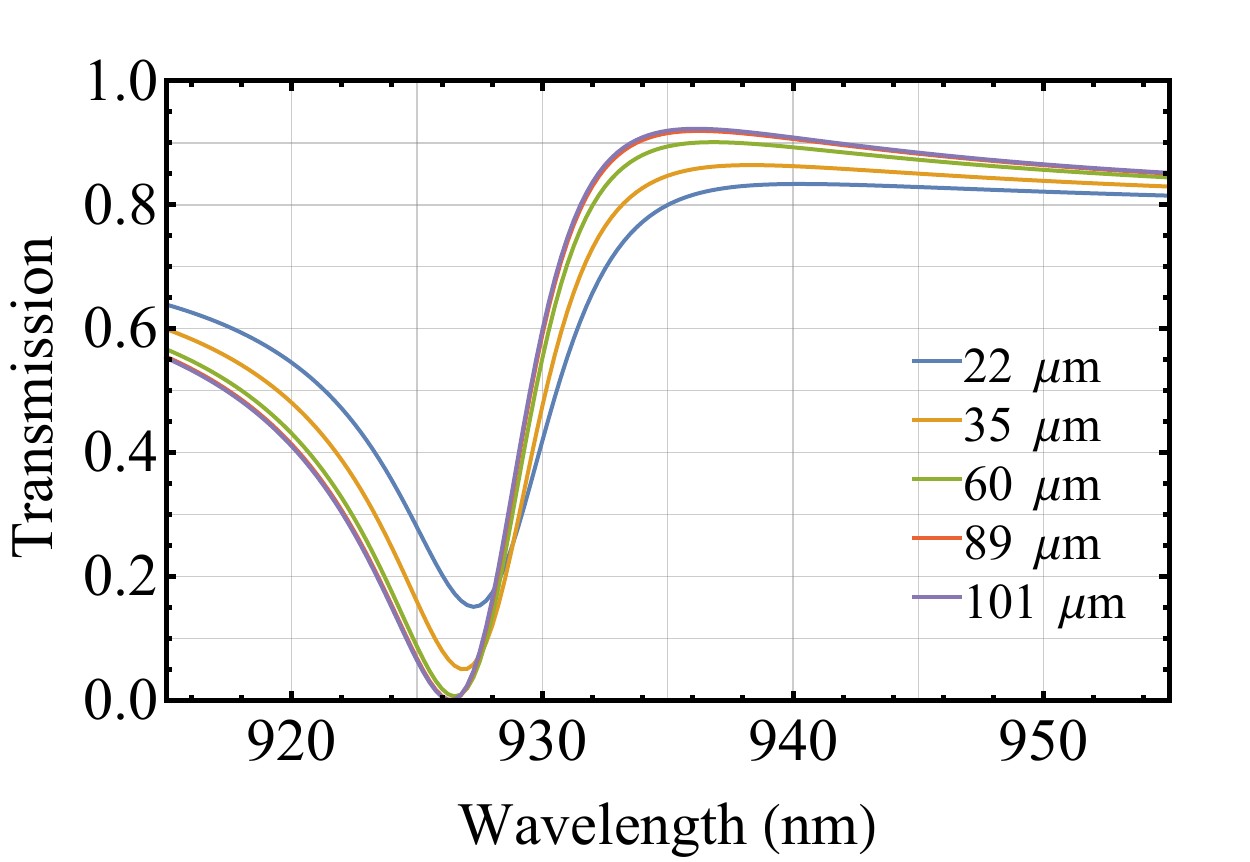}\includegraphics[width=0.85\columnwidth]{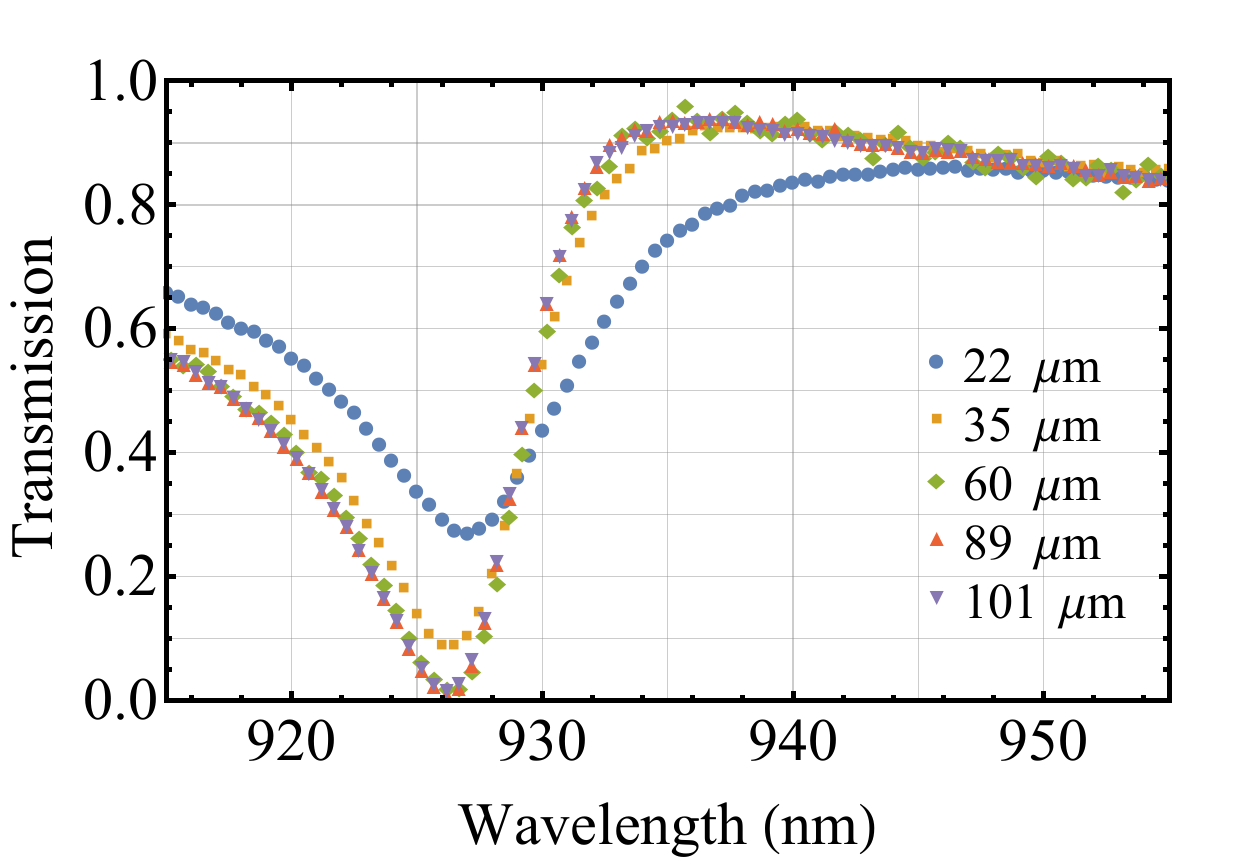}
\caption{Sample C. Simulated transmission spectra including collimation effects only (top left), finite-size effects only (top right) and both effects (bottom left), as well as the corresponding experimental spectra (bottom right).}
\label{fig:C4}
\end{figure*}

\begin{figure*}
\includegraphics[width=0.85\columnwidth]{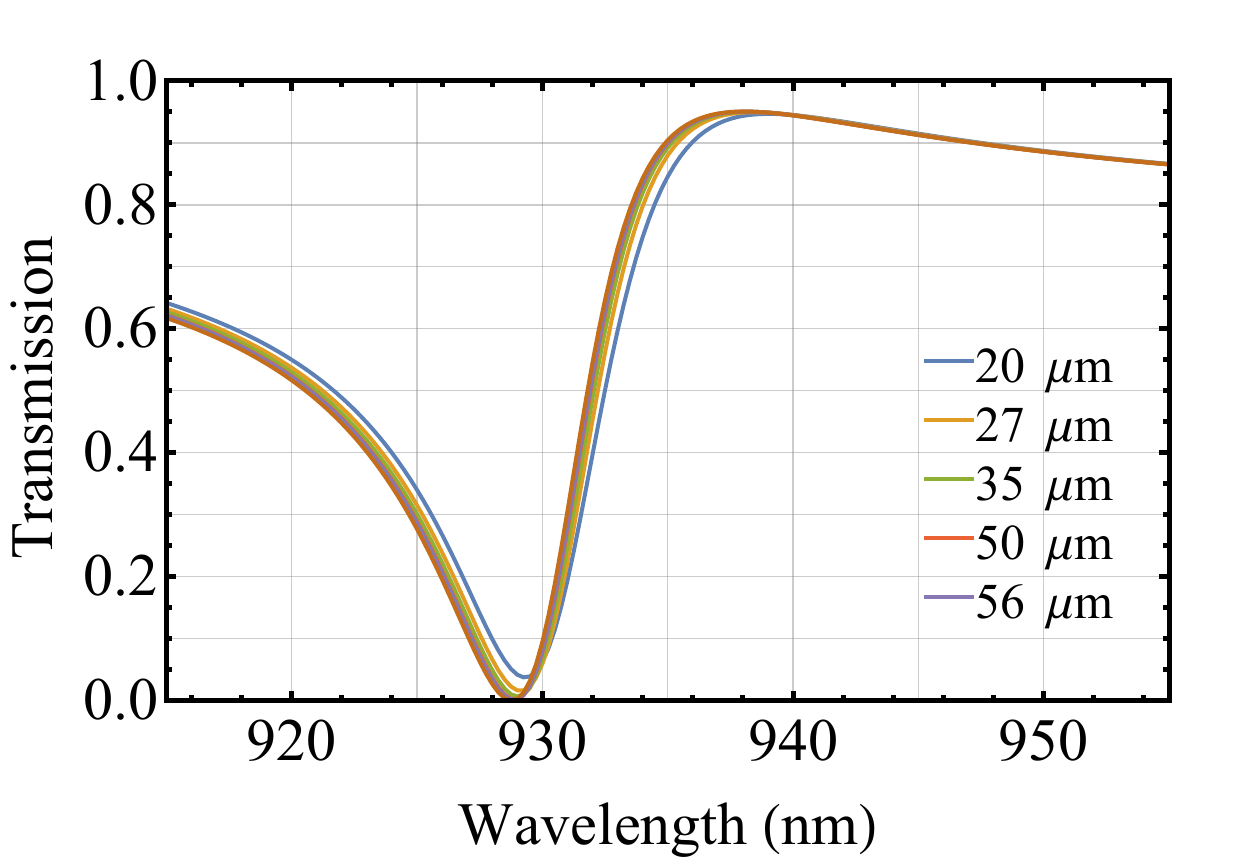}\includegraphics[width=0.85\columnwidth]{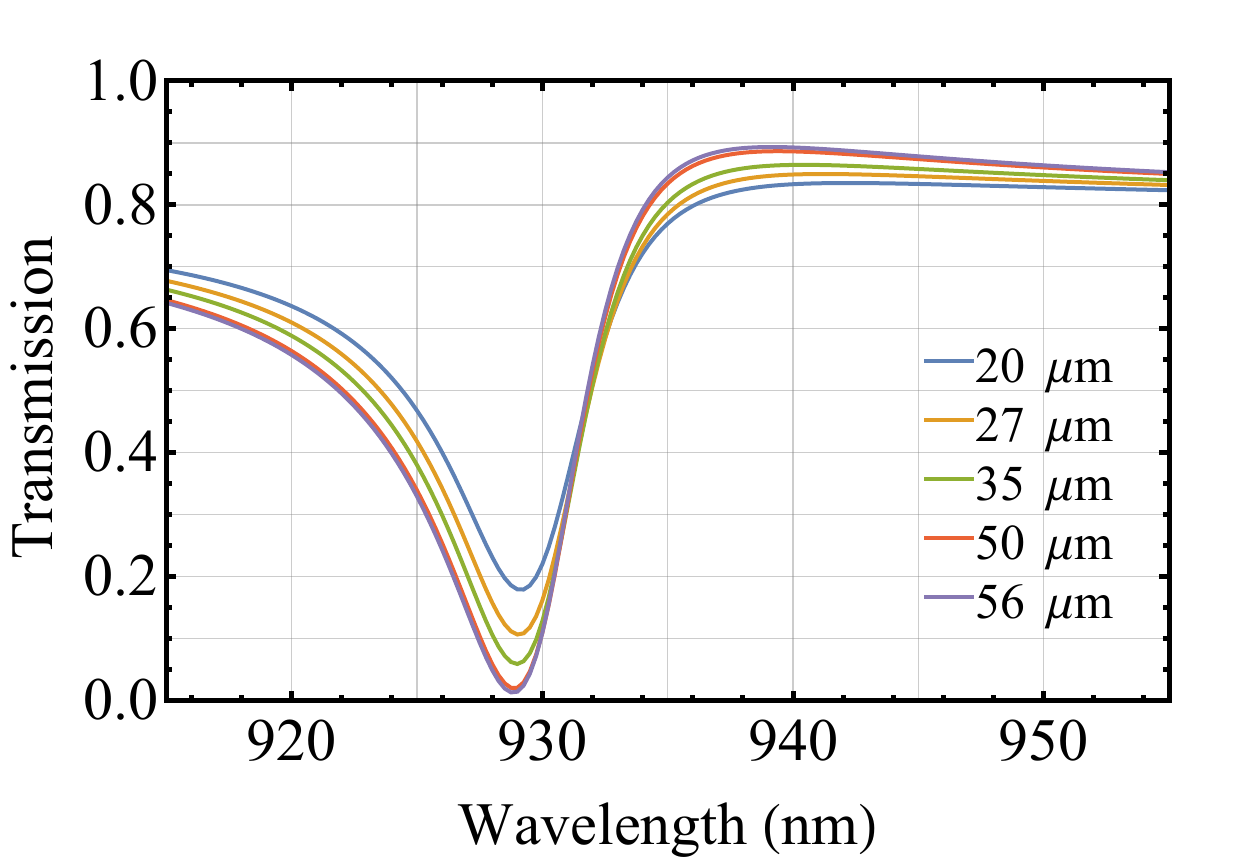}\\
\includegraphics[width=0.85\columnwidth]{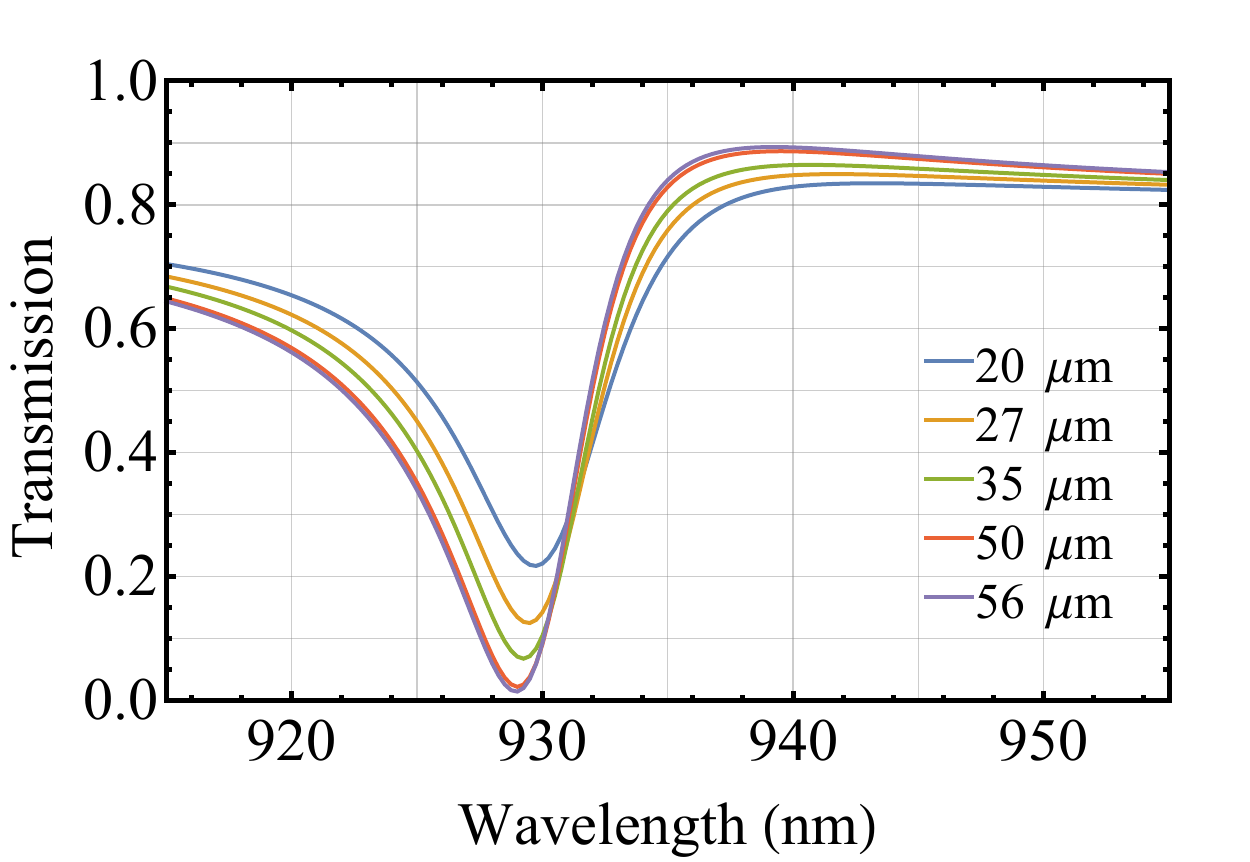}\includegraphics[width=0.85\columnwidth]{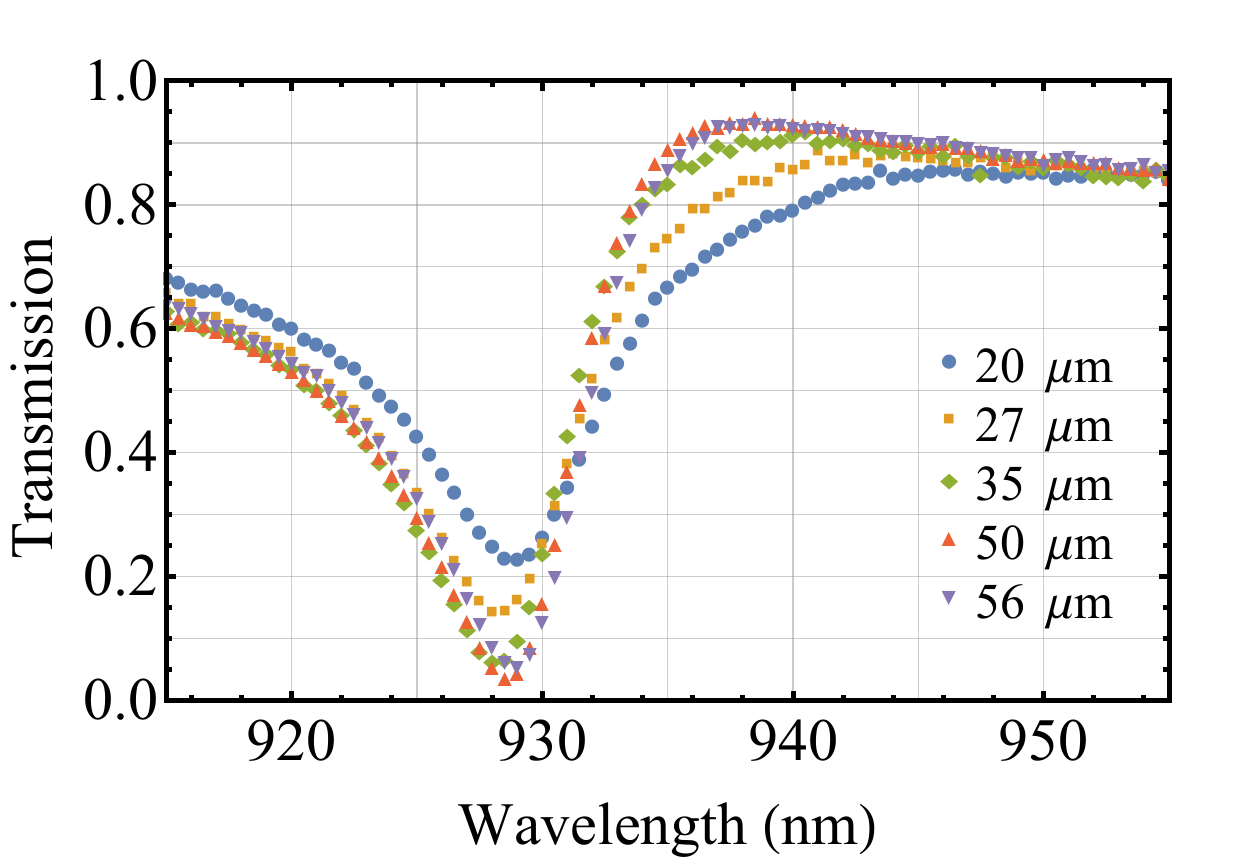}
\caption{Sample D. Simulated transmission spectra including collimation effects only (top left), finite-size effects only (top right) and both effects (bottom left), as well as the corresponding experimental spectra (bottom right).}
\label{fig:D4}
\end{figure*}

\begin{figure*}
\includegraphics[width=0.85\columnwidth]{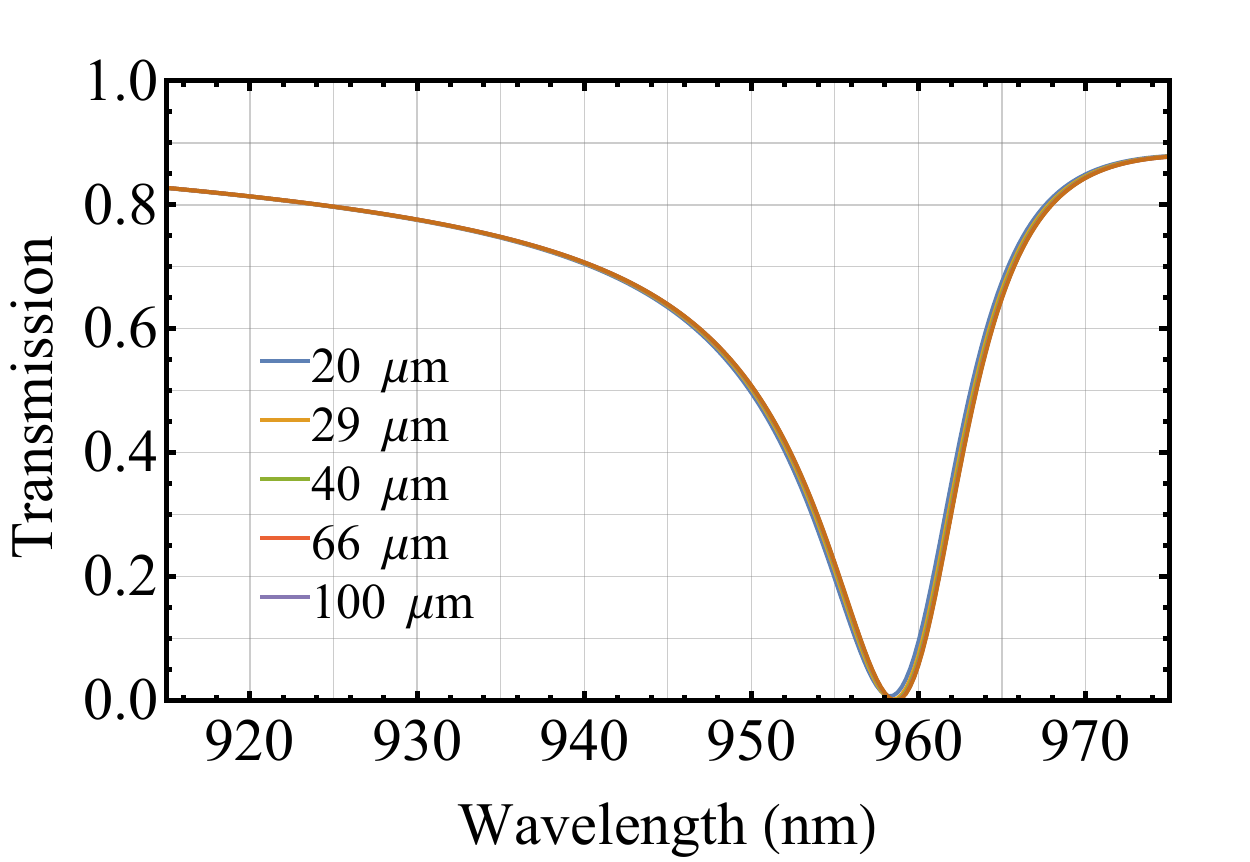}\includegraphics[width=0.85\columnwidth]{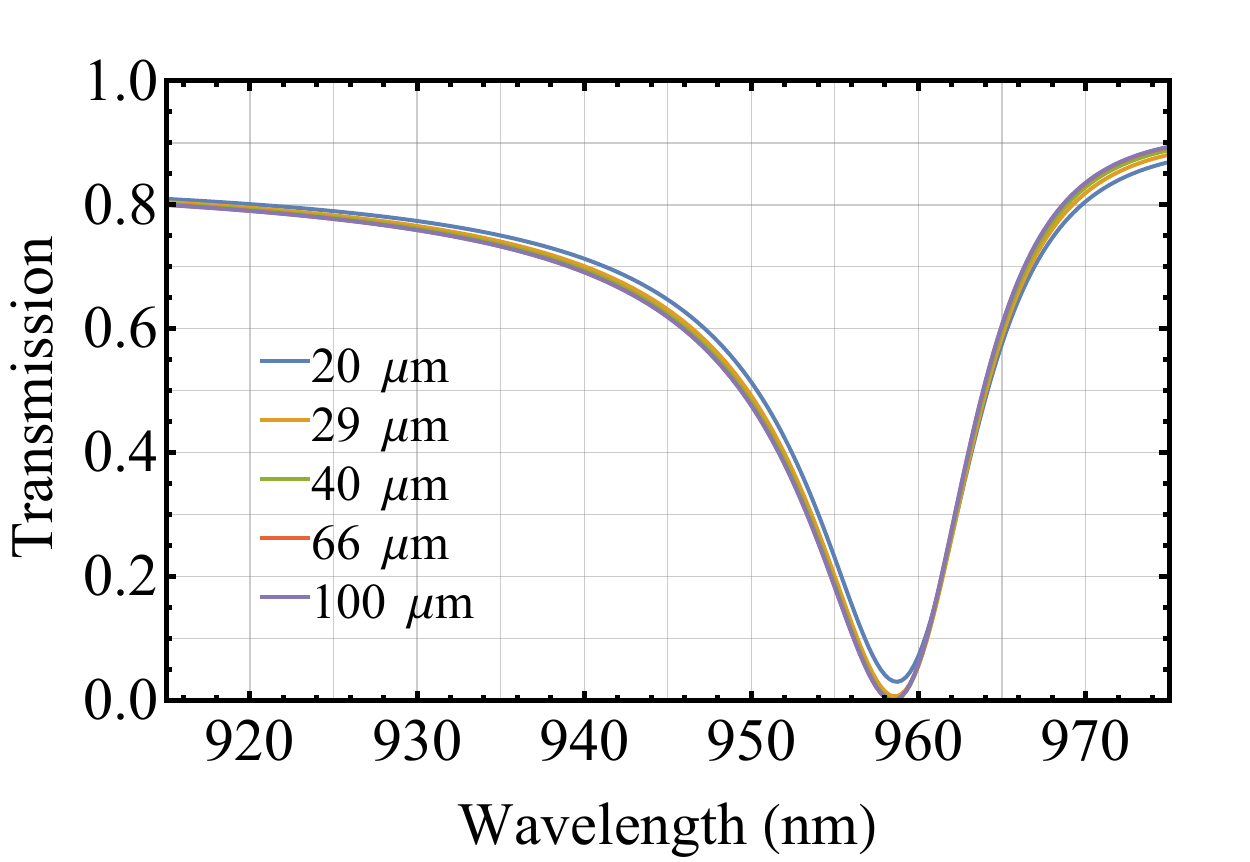}\\
\includegraphics[width=0.85\columnwidth]{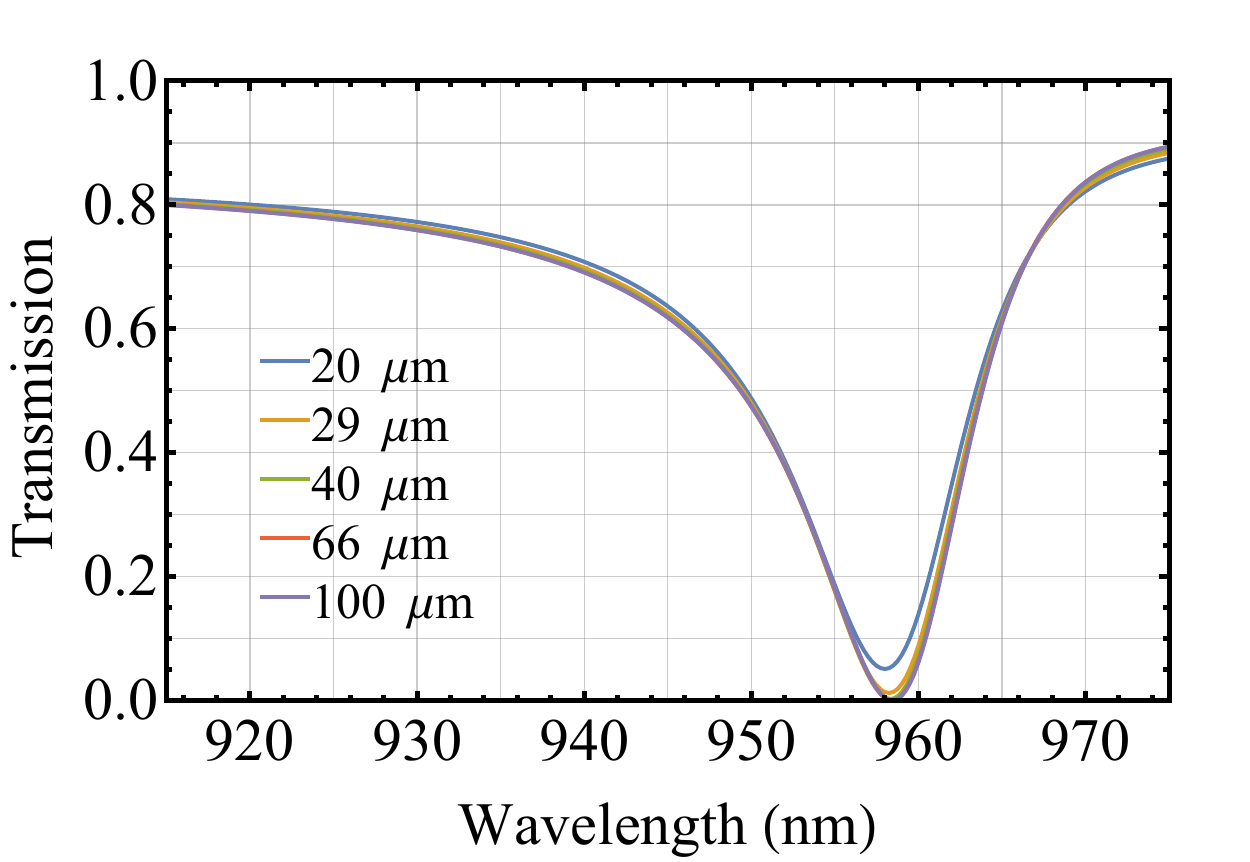}\includegraphics[width=0.85\columnwidth]{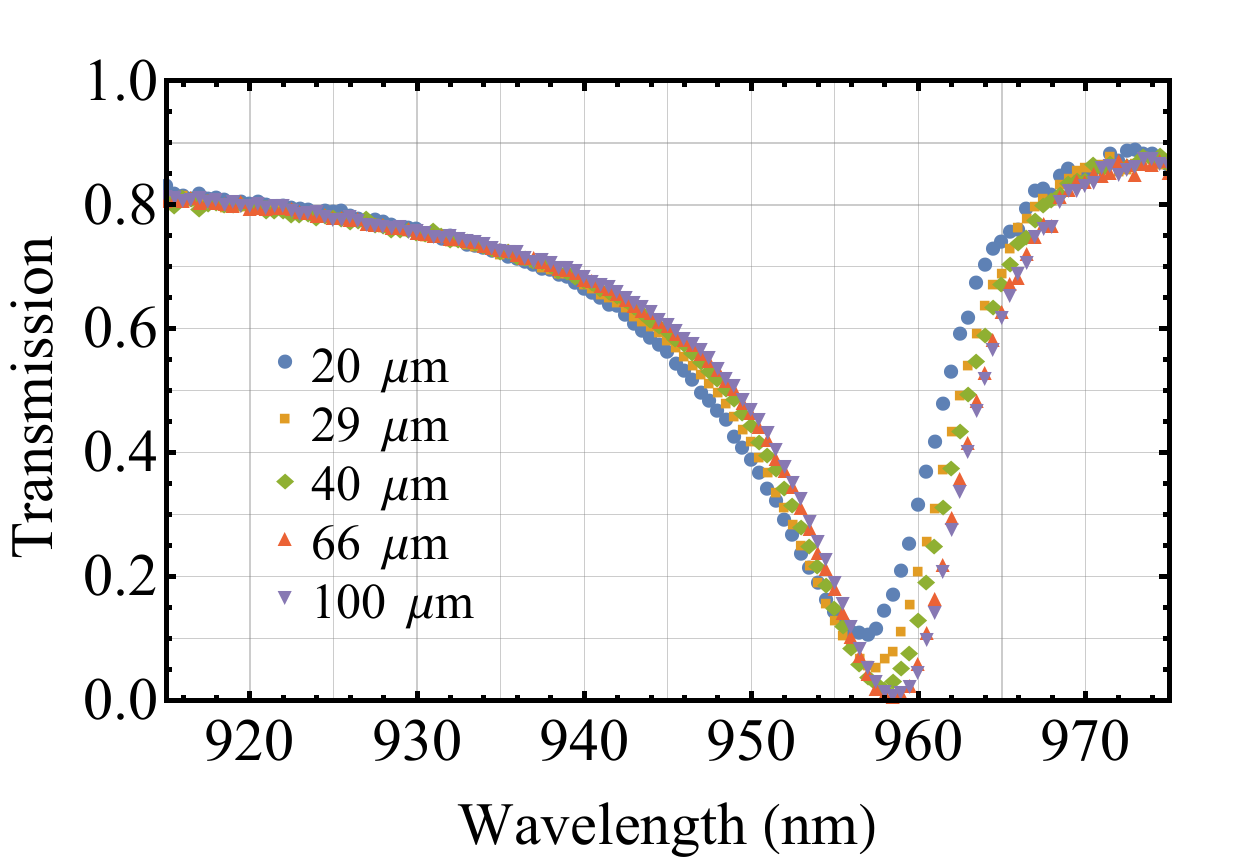}
\caption{Sample E. Simulated transmission spectra including collimation effects only (top left), finite-size effects only (top right) and both effects (bottom left), as well as the corresponding experimental spectra (bottom right).}
\label{fig:E4}
\end{figure*}

Similar analyses were performed for the four other gratings and the resulting spectra are reported in Figs.~\ref{fig:B4}-\ref{fig:E4}, while the resulting model/fit parameters are given in Table~\ref{tab:tab2}.

\begin{table}[h]
\caption{\label{tab:parameters}Analysis results for samples A-E.}
\label{tab:tab2}
\begin{ruledtabular}
\begin{tabular}{cccccc}
Sample & A & B & C & D & E \\\hline
$\lambda_0$ [nm] & 941.8 & 948.3 & 926.3 & 928.7 & 958.3\\
$\lambda_2$ [nm] & 925.4 & 918.6 & 900.1 & 900.0 & 1002.0\\
$\nu$ [nm$^{-2}$] & $2.9\,10^{-5}$ & $2.2\, 10^{-5}$ & $2.4\, 10^{-5}$ & $2.8\, 10^{-5}$ & $2.4\, 10^{-5}$\\\hline
$\nu_\theta$ [nm$^{-2}$] & $1.6\,10^{-3}$ & $6.8\,10^{-4}$ & $8.2\,10^{-4}$ & $8.7\,10^{-4}$ & $-5.5\,10^{-4}$\\\hline
$\delta_\lambda$ [nm] & 2.9 & 2.6 & 3.6 & 3.3 & 5.8\\
$l$ [$\mu$m] & 3.0 & 3.4 & 2.4 & 2.3 & 1.1\\
$\eta$ [\%] & 8.2 & 7.8 & 9.0 & 8.3 & 8.4\\\hline
$A_\textrm{w}$ [$\mu$m] & 37.0 & 43.1 & 26.7 & 29.2 & 12.6
\end{tabular}
\end{ruledtabular}
\end{table}

The analysis corroborates well the qualitative conclusions drawn from the bare experimental observations of Sec.~\ref{sec:exp_results}. It can for instance be seen that the higher value of $\nu_\theta$ for sample A than for sample B suggests a higher sensitivity to collimation effects, while its comparatively lower $A_\textrm{w}$ indicates a lower sensitivity to finite-size effects. That explains why both SWGs overall perform similarly, in spite of the noticeable difference in finger depth. Samples C and D are observed to have similar sensitivities to both effects, but their larger grating coupling strength and lower modal dephasing make them comparatively less sensitive to finite-size effects than samples A and B. The "TE grating" (sample E) possesses the lowest $\nu_\theta$ and $A_\textrm{w}$ values, due to its larger angular width and reduced modal dephasing, and, as such, displays the lowest sensitivity to both collimation and finite-size effects.

The remaining small deviations between the experimental and predicted spectra could in principle be due to inhomogeneities of the grating structures, absorption or the structural deformation of the grating area resulting from the patterning~\cite{Darki2021}. While effects due to inhomogeneities and absorption can be estimated and are negligible for these films, the effect of the deformation is more complex to predict and would require full three-dimensional finite-element simulations of the structures. Other possible sources of discrepancy are the small variations in the waist size ($<5$\%) and position ($<0.2$ mm) with the wavelength in the range of interest, as well as the uncertainty in positioning the grating at the focus and aligning it at normal incidence.

\section{Conclusion}\label{sec:conclusion}

A detailed experimental and theoretical investigation of finite-size and collimation effects in five suspended resonant guided-mode gratings patterned on ultrathin suspended silicon nitride films was carried out. While possessing similar parameters these gratings differ in terms of either finger depth, period og size of the patterned area, yielding different sensitivities with respect to finite-size and collimation effects. High reflectivity Fano resonances in the range 915-975 nm are experimentally observed for various focusing of the incidence Gaussian beam. A phenomenological coupled-mode model was put forward in order to quantify the relative magnitude of these effects on these resonances. This model is based on a one-dimensional angular spectrum representation of the incident field and a phenomelonogical extension of the coupled-mode model of Bykov et al.~\cite{Bykov2015} using a waveguide interference model put forward by Jacob et al.~\cite{Jacob2000,Jacob2001}. The predictions of the model are observed to be in overall very good agreement with the experimental measurements. Such a simple model could thus be used to optimize the design of such patterned suspended films for applications with focusing constraints, e.g. for cavity optomechanics, lasing or integrated optomechanical microcavities.


\section*{Funding}
Independent Research Fund Denmark.

\section*{Disclosures} 
The authors declare no conflicts of interest.


\end{document}